\def\rn{}
\def\nn#1 #2{#2. #1}				% Name with 1 initial
\def\nnn#1 #2 #3{#2. #3. #1}			% Name with 2 initials
\def\nnnn#1 #2 #3 #4{#2. #3. #4. #1}		% Name with 3 initials
\def\nnnnn#1 #2 #3 #4 #5{#2. #3. #4. #5. #1}	% Name with 4 initials
\def\dualand{ and\hbox{ }}				
\def\multiand{ and,\hbox{ }}				
\def\rf#1;#2;#3;#4;#5 {{\frenchspacing\par\rn#1, #3 {\bf #4}, #5 (#2). \par}}
\def\rfbook#1;#2;#3;#4;#5 {{\frenchspacing\par\rn#1, {\it #3} (#5, #4, #2).\par}}
\def\rfprep#1;#2;#3 {{\par\frenchspacing\rn#1, #3 (#2).\par}}
\def\Mpc{{\rm Mpc}}

\def\expec#1{\langle#1\rangle}

\def\etal{{\frenchspacing\it et al.}}
\def\ie{{\frenchspacing\it i.e.}}
\def\eg{{\frenchspacing\it e.g.}}
\def\etc{{\frenchspacing\it etc.}}

\def\crr{\cr\noalign{\vskip 4pt}}

\def\beq#1{\begin{equation}\label{#1}}
\def\eeq{\end{equation}}
\def\beqa#1{\begin{eqnarray}\label{#1}}
\def\eeqa{\end{eqnarray}}
\def\eq#1{equation~(\ref{#1})}
\def\Eq#1{Equation~(\ref{#1})}
\def\eqn#1{~(\ref{#1})}

\def\fig#1{Figure~\ref{#1}}
\def\Fig#1{Figure~\ref{#1}}

\def\sec#1{Section~\ref{#1}}

\def\spose#1{\hbox to 0pt{#1\hss}}
\def\simlt{\mathrel{\spose{\lower 3pt\hbox{$\mathchar"218$}}
     \raise 2.0pt\hbox{$\mathchar"13C$}}}
\def\simgt{\mathrel{\spose{\lower 3pt\hbox{$\mathchar"218$}}
     \raise 2.0pt\hbox{$\mathchar"13E$}}}
\def\simpropto{\mathrel{\spose{\lower 3pt\hbox{$\mathchar"218$}}
     \raise 2.0pt\hbox{$\propto$}}}

\def\ed{\end{document}}

\def\Ot{\Omega_0}

\def\Ol{\Omega_\Lambda}
\def\Om{\Omega_{\rm m}}

\def\Geff{G_{\rm eff}}
\def\Gbare{G_{\rm bare}}

\def\k{{\bf k}}

\def\C{{\bf C}}

\def\F{{\bf F}}

\def\l{\ell}

\def\nth{n^{th}}

\def\deltahat{\widehat{\delta}}

\def\flum{d_{lum}}
\def\fang{d_{ang}}
\def\fage{d_{age}}

\def\fest{{\tilde f}}

\def\rhoeff{\rho_{\rm eff}}
\def\rhoeffz{\rho_{\rm eff}(z)}

\documentstyle[prl,aps,epsf]{revtex}
\begin{document}
\twocolumn[\hsize\textwidth\columnwidth\hsize\csname@twocolumnfalse\endcsname

%%%%%%%%%%%%%%%%%%%%%%%%%%%%%

\preprint{IASSNS-AST 97/666}

\title{Measuring the metric: 
a 
parametrized 
post-Friedmanian approach to the cosmic dark energy problem}
% Hidden energy or hidden assumptions?

\author{Max Tegmark}

\address{Dept. of Physics, Univ. of Pennsylvania, 
Philadelphia, PA 19104; max@physics.upenn.edu}

%\date{June 19, 1997}
%\date{Submitted 30 April 2000; published 17 November}
\date{Submitted to Phys. Rev. D Jan 22 2001, accepted Aug 16 2002}

\maketitle

\begin{abstract}
We argue for a ``parametrized post-Friedmanian'' approach 
to linear cosmology, where the history of
expansion and perturbation growth is measured without
assuming that the Einstein Field Equations hold.
As an illustration, a model-independent analysis of 
92 type Ia supernovae demonstrates that the
curve giving the expansion history has the wrong shape
to be explained without some form of dark energy or modified
gravity. We discuss how upcoming lensing, galaxy clustering,
cosmic microwave background and Ly$\alpha$ forest observations can be combined 
to pursue this program, and forecast 
the accuracy that the proposed SNAP satellite 
can attain.
\end{abstract}


] % Must end \twocolumn command here,  or disaster occurs, for bizarre reasons.

%%%%%%%%%%%%%%%%%%%%%%%%%%%%%%%%%%%%%%%%%%%%%%%
%\makeatletter
%\global\@specialpagefalse
%%\def\@oddhead{Tegmark\hfill Measuring CMB power spectra}
%%\let\@evenhead\@oddhead
%\def\@oddfoot{
%\ifnum\c@page>1
%  \reset@font\rm\hfill \thepage\hfill
%\fi
%\ifnum\c@page=1
%Published in {\it\frenchspacing Phys. Rev. Lett.} {\bf 79}, 3806 (1997).
%Available in color from 
%{\it h t t p://www.sns.ias.edu/$\tilde{~}$max/galfisher.html} \hfill\\
%\fi
%} \let\@evenfoot\@oddfoot
%\makeatother

\section{INTRODUCTION}

Modern cosmology is in a somewhat equivocal
% = Open to two or more interpretations, tvetydig
state of affairs.
The good news is that a recent avalanche of high-quality
data are well fit by an emerging 
``standard model'' whose roughly ten free parameters are being 
constrained with increasing precision 
\cite{Lange00,boompa,Bambi00,Bridle00,observables,Jaffe00,Kinney01,Durrer00,concordance,consistent,Efstathiou02}. 
The bad news is that this emerging model is more complicated
than anticipated. There is not one kind of invisible substance, but three: cold dark matter (CDM) to explain clustering,
dark energy to close the Universe and a smidgeon of massive neutrinos
\cite{Scholberg99}
that may well be too small to be cosmologically important.
Moreover, problems involving small-scale clustering have triggered
increasingly complicated models for the dark matter
--- self-interacting CDM 
\cite{Carlson92,deLaix95,Spergel00,Hogan00,Hannestad00,Burkert00,Firmani00,Mo00,Kochanek00,Yoshida00}, 
annihilating CDM \cite{Kaplinghat00}, 
warm dark matter \cite{Bonometto85,Schaeffer88,Colombi96,Colin00,Narayanan00,Bode00,Carlson92,Hogan00,Hannestad00}, 
fuzzy dark matter \cite{Hu00}
and fluid dark matter \cite{Peebles00a},
% non-interacting dark matter \cite{PeeblesVilenkin00}
to mention a few --- and a large number of 
dark energy models ({\eg} \cite{Zlatev,SahniStarobinski00}) have appeared where Einstein's 
single parameter $\Lambda$ is replaced by a ``quintessence''
field that varies temporally and perhaps spatially.

This perceived profusion of bells and whistles has 
caused unease among some cosmologists 
\cite{Peebles99,Peebles00,Sellwood00,Disney00,McGaugh00} 
and prompted concern that 
these complicated dark matter flavors constitute a
modern form of epicycles.
There have been numerous suggestions that, 
just as in the days of Ptolemy, the apparent complications
can be eliminated by modifying the laws of 
gravity \cite{Milgrom83,Milgrom98,Damour99,Mannheim00,Boisseau00,Esposito00,Gaztanaga00,Binetruy00,Uzan00,Hwang02,Behnke02}
% astro-ph/0102005 too?
For instance, 
there are scalar-tensor theories
that can reproduce the observed accelerating
Universe without invoking a cosmological constant
\cite{Boisseau00,Esposito00,Gaztanaga00}.

It will undoubtedly take time to settle these issues.
In the interim, however, it is desirable to quote
cosmological measurements in a language that is 
fairly theory-independent. This is the topic of the present paper,
delimited to gravity in the linear regime, on large scales.

%%%%%%%%%%%%%%%%%%%%%%%%%%%%%%%%%%%%%%%%%%%%%%%%%%%%%%%%%
\begin{figure}[phbt]
%\vskip-1.2cm
\centerline{{\vbox{\epsfxsize=9.5cm\epsfbox{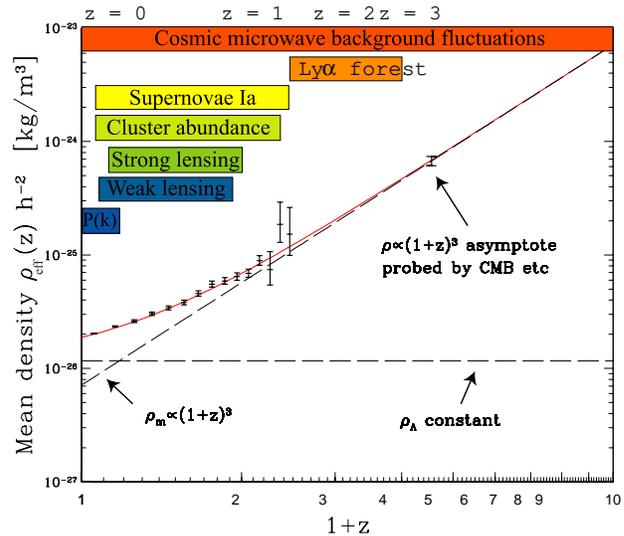}}}}
\vskip-8cm
%\smallskip
%\centerline{{\vbox{\epsfxsize=11cm\epsfbox{fig1.ps}}}}
\label{rhoFig}
\caption{
Solid curve shows a currently popular model
for the evolution of the effective cosmic mean density $\rhoeff(z)\propto H(z)^2$.
This curve uniquely characterizes the spacetime metric to zeroth order.
The horizontal bars indicate the rough
redshift ranges over which the various cosmological probes discussed
are expected to constrain this function.
If the Friedmann equation is correct, then  $\rhoeff(z)=\rho(z)$.
Since the redshift scalings of all density contributions except that of
dark energy are believed to be straight lines with known slopes
in this plot (power laws),
combining into a a simple quartic polynomial,
an estimate of the
dark energy density $\rho_X(z)$ can be readily extracted from 
this curve. 
Specifically, 
$\rho\propto (1+z)^4$ for the Cosmic Microwave Background (CMB), 
$\rho\propto (1+z)^3$ for nonrelativistic matter (like baryons and CDM),
$\rho\propto (1+z)^2$ for spatial curvature,
$\rho\propto (1+z)^0$ for a cosmological constant
and $\rho\propto (1+z)^{3(1+w)}$ for quintessence with a constant equation
of state $w$. Error bars are for our SNAP SN 1a simulation.
}
\label{rzFig}
\end{figure}
%%%%%%%%%%%%%%%%%%%%%%%%%%%%%%%%%%%%%%%%%%%%%%%%%%%%%%%%%

\subsection{Parametrizing our ignorance with two functions}

There is general consensus that space-time is well described by a metric,
at least on non-microscopic scales, so competing theories of gravity 
differ in their predictions for how the metric evolves with time and responds
to the presence of matter. General Relativity (GR) makes specific
predictions in the form of the Einstein Field Equations.
The parametrized post-Newtonian approximation \cite{WillBook,Will98}
has been highly successful in describing possible departures from 
GR in the solar system, and the
parametrized post-Keplerian approximation \cite{WillBook,Will98}
has done the same for 
for binary orbits, notably binary pulsars. 
Although GR has so far passed all experimental tests
with flying colors, these tests tend to probe only the present epoch.
Broad classes of theories have been shown to evolve into GR at
late times, so substantial departures from the Einstein field
equations are still possible in cosmology.

Assuming merely the cosmological principle 
(that our position is not special), it follows directly from the observed
near-isotropy of the cosmic microwave background (CMB) that the 
Universe is approximately isotropic and homogeneous. 
This means that the large-scale metric consists of small perturbations on a 
Friedman-Robertson-Walker (FRW) background \cite{Weinberg72}:
\beq{FRWeq}
ds^2 = dt^2 - a(t)^2\left[{dr\over 1-kr^2} - r^2 d\theta^2 - \sin^2\theta d\phi^2\right]
%ds^2 = dt^2 - a(t)^2\left[{dr\over 1-kr^2} - r^2 d\Omega^2]
\eeq
for some function $a(t)$ that describes the expansion history of the Universe.
$k=-1$, $0$ or $1$ for an open, flat and closed Universe, respectively.
If GR is correct, then $a$ evolves according to the Friedman equation \cite{Weinberg72}
\beq{FriedmanEq}
H^2\equiv \left({\dot a\over a}\right)^2 = {8\pi G\over 3}\rho,
\eeq
where we define the mean matter density $\rho(z)$ to include a curvature
contribution $\rho_k\equiv -{3kc^2\over 8\pi G a^2}$.
Other theories make different predictions.
As summarized in \fig{rzFig} and reviewed in the next section, the function 
$H(z)$ is directly measurable is a variety of ways that make no assumptions
about dark matter properties or gravitational field equations, but rely
on spacetime geometry alone.
     
Let us now turn to the evolution of perturbations.
The evolution equations are by definition linear to first order,
regardless what underlying theory of gravity is being linearized.
Assuming that this theory is well approximated by 
partial differential equations on large scales, 
these decouple into ordinary differential 
equations for each Fourier mode,
since we are perturbing around a homogeneous metric and
derivatives become local in Fourier space.
In short, the key observable predicted by theory is
the linear growth factor $g(z,k)$, conventionally defined
so that a density perturbation $\deltahat(\k)$ with wave vector $\k$ 
evolves with redshift as 
\beq{gDefEq}
\delta(z,\k) = {g(z,k)\over (1+z)} \delta(0;k).
\eeq

A wide variety of upcoming observations will
probe $g(z,k)$, including galaxy and Lyman $\alpha$ forest clustering,
the cosmic microwave background,
gravitational lensing and type Ia supernovae.
When multiple matter components are present, 
there will of course be one growth factor for each one,
whose evolution couple. For photons and neutrinos,
the full phase-space distribution is relevant to the dynamics.
On large scales, a broad class of models make the simple
prediction that $g(z,k)$ is independent of $k$.

\subsection{Is matter the matter?}

It is obviously premature to guess as to whether
upcoming data will favor dark matter or modified gravity.
However, it is worth noting that the distinction between
these two cases is in fact somewhat blurry.
As pointed out by Eddington \cite{Eddington}, 
we can always choose to {\it define} matter as that
which equals $(8\pi G)^{-1}$ times the Einstein tensor, 
in which case the Einstein field equations 
$G_{\mu\nu}=8\pi G T_{\mu\nu}$ become satisfied 
by definition. This ``matter'' $T_{\mu\nu}$ 
would have very strange properties
for a generic metric, so the predictive power of GR arises from
the fact that observed matter has simple, formalizable properties.

For many modifications of GR, the so defined $T_{\mu\nu}$ is in fact rather regular.
Indeed, in the familiar case of adding a cosmological constant,
it is largely a matter of taste whether 
to call it modified gravity or dark energy, 
corresponding merely to whether we insert  it on the left or right
hand side of the Einstein equations.
In the currently popular class of GR generalizations known as
scalar-tensor theories \cite{Weinberg72,WillBook,Esposito00}, 
of which Brans-Dicke theory 
is a special case, we can play the same trick: the Friedman equation\eqn{FriedmanEq}
remains valid in the conventional (Jordan) 
frame \cite{Esposito00} 
if we augment the density $\rho$ with the extra 
``dark energy'' term 
\beq{ScalarTensorEq}
\rho_x = \left({1\over\zeta F}-1\right)\rho_m + 
{1\over 8\pi\Geff F}\left[U + {Z{\dot\Phi}^2\over 2} - 3H\dot{F}\right],
\eeq
where the theory is specified by the free functions $F(\Phi)$, $U(\Phi)$ and $Z(\Phi)$ 
of the dilaton field $\Phi$, $\rho_m$ is the density of conventional matter, and
\beq{etaDefEq}
\zeta\equiv{\Geff\over\Gbare}={1\over F}\left({2ZF+4(F')^2\over 2ZF+3(F')^2}\right)
\eeq
is the factor by which $\Geff$, the effective small-scale Newton's constant 
measured in laboratory experiments today, differs from the ``bare'' 
one $\Gbare$ which is truly constant.
%Here $F'$ is the derivative of $F$ with respect to the dilaton field $\Phi$ upon
%which it depends.
% Here and throughout, dots and primes denote derivatives with respect to time and 
% redshift, respectively; $d/dz = -[H(z)/(1+z)]d/dt$.

As another  example, consider a flat toy model where 
\eq{FriedmanEq} is replaced by 
\beq{ToyFriedmanEq}
H^2 = {8\pi G\over 3}{\rho^2\over\rho_*}
\eeq
for some constant $\rho_*$, \ie, a model where $\rho\propto H$ rather than
$\rho\propto H^2$. An observer assuming that the Friedman equation \eq{FriedmanEq}
is valid would then conclude that there is a ``dark energy'' density $\rho_x$ given by
\beq{ToyDarkEnergyEq}
\rho_x \equiv {3H^2\over 8\pi G} - \rho = {\rho^2\over\rho_*}-\rho
= {\rho_0^2\over\rho_*}(1+z)^6 - \rho_0(1+z)^3,
\eeq
a quantity which goes negative for small $\rho$. 
Here $\rho_0$ denotes the current density of baryonic and dark matter.

The modified behavior of linear perturbations in non-GR theories
can clearly be attributed to 
new ``matter'' as well. Indeed, it has recently been shown \cite{HuGDM}
that the linearized properties of very general types of dark matter can 
be conveniently characterized by an effective sound speed and an effective viscosity.

In light of this theory/matter ambiguity, 
we propose to use the Friedman equation as a {\it definition} 
of an effective mean density 
\beq{rhoDefEq}
\rhoeff(z)\equiv {3H(z)^2\over 8\pi\Geff}.
\eeq
Thus  $\rhoeff(z)=\rho(z)$, the true density, if the Friedmann equation is correct.
As illustrated in \fig{rhoFig}, expressing observational results in terms of 
$\rhoeff(z)$ rather than $H(z)$ has the advantage 
of directly visualizing the dark matter problem.
The contribution to $\rho(z)$ from vacuum density is of course constant.
Curvature density scales as $(1+z)^2$,
matter density as $(1+z)^3$, 
radiation density as $(1+z)^4$, and a ``quintessence''.
component with constant equation of state $w$
as $(1+z)^{3(1+w)}$.

A large body of work has been performed during the last few years
on constraints and forecasts for more specific models. Cast in our
present notation, a particularly useful and popular
parametrization of $\rho(z)$ has been 
\cite{Garnavich98,Starobinski98,Efstathiou99,Perlmutter99,Cooray99,Huterer99,Huterer00,Newman00,Lima00,Saini00,Bean01,Kujat01,Maor02a,Maor02b,Weller02,Wasserman02,Spergel02,Peebles02,Friemann02,Linder02}
\beq{rhoEq2}
\rho(z) = 
%\rho^0_\gamma(1+z)^4 + %
\rho^0_m(1+z)^3
+ \rho^0_\kappa(1+z)^2 + \rho^0_X (1+z)^{3(1+w)}
\eeq
for $z\ll 10^3$,
where
%$\rho^0_\gamma$, 
$\rho^0_m$, $\rho^0_\kappa$
and $\rho_X^0$ are the present densities of 
matter, curvature and dark energy, respectively, 
and the dark energy equation of state $w$ may in turn vary with redshift.
In this paper, we treat $\rhoeff(z)$ rather than $w(z)$ as the 
free function for two reasons:
\begin{enumerate}
\item It can be estimated from the data without assumptions 
about gravity (the Einstein field equations) 
or about the dark energy (which may be out of equilibrium and not have any
simple equation of state uniquely determining its pressure from its 
density).
\item It is more directly related to observed data since, as will be elaborated
below, it is given by the first rather than the second derivative
of a measured function.
\end{enumerate}
A similar choice was made in \cite{Wang01} as well.

Much progress has also been made on a reconstruction program, aiming to 
reconstruct the microphysics of specific theories from data.
The present work complements this approach, since such 
reconstructions of, \eg, the quintessence potential 
\cite{Starobinski98,Saini00,Huterer00,Maor02a,Maor02b}
or the functions specifying scalar-tensor gravity \cite{Boisseau00}
can be performed with $\rhoeff(z)$ and $g(z,k)$ as a starting point.

The rest of this paper is focused on 
the 0th order function $\rhoeff(z)$, with the 1st order function
$g(z,k)$ studied in a companion paper \cite{pwindows}.
In the next section, we review techniques for measuring 
$\rhoeff(z)$, compute the constraints from current SN 1a data and make 
forecasts for the proposed SNAP satellite.
We discuss our conclusions in \sec{DiscussionSec}.

%\subsection{Leftovers}
%
%KINEMATICS AS OPPOSED TO DYNAMICS.
%
%CITE SOME OTHER NON-STANDARD GRAVITIES TOO:
%large extra dimensions, string theory, $R^2$-theories, etc.

\section{Measuring $\rhoeffz$}

The bars at the top of \fig{rhoFig} show the redshift ranges over which various 
types of observations are sensitive to $\rhoeff(z)$. We will first 
review them briefly in turn, then return the SN 1a in greater detail.

\subsection{Space-time geometric tests}

All cosmological tests based on luminosity, angular size and age
(see {\eg} \cite{Weinberg72,Peebles93})
can be described as noisy measurements of some quantities
$x_n$ at redshifts $z_n$, $n=1,...,N$, modeled as
\beq{xEq}
x_n = a\ln d(z) + b_n + \varepsilon_n,
\eeq
where $a$ and $b$ are constants independent of $\rhoeff(z)$, 
the function $d$ incorporates the effects of cosmology 
and $\varepsilon_n$ is a random term with zero
mean ($\expec{\varepsilon_n}=0$)
including all sources of measurement error.
As briefly review below, these observables all 
probe different weighted averages of the quantity $\rhoeff(z)^{-1/2}$.

For luminosity tests like SN Ia, $x_n$ is the observed magnitude 
of the $\nth$ object and $d$ is the dimensionless luminosity distance \cite{Weinberg72}:
\beq{LumDistEq}
\flum  = (1+z){S(\kappa\eta)\over\kappa},\quad
\eta(z) = H_0\int_0^z{dz'\over H(z')},
\eeq
where the $\kappa\equiv\sqrt{|1-\Ot|}$ is the spatial curvature,
and $\Ot\equiv 8\pi G\rhoeff(0)/3H^2$ is the fraction of critical density
contributed by $\rhoeff$ today. $S(x)\equiv\sinh x$, $x$ or $\sin x$
when $k=-1$, $0$ or $1$, respectively.
From the definition of magnitudes, $a=5/\ln 10$ and
$b_n = 25 + 5\lg[H_0^{-1} c/1\Mpc] + M_n$, where $M_n$ is the (hopefully known)
absolute magnitude of the $\nth$ object.
The errors $\varepsilon_n$ include 
errors in extinction correction and intrinsic scatter in the ``standard candle''
luminosity.

For tests involving the observed angular sizes $\theta_n$ of objects
at redshifts $z_1,...,z_N$ with (hopefully known) linear sizes
$D_1,...,D_N$, 
we can take $d$ to be the dimensionless angular size 
distance \cite{Weinberg72}: $\fang=\flum/(1+z)^2$.
This gives $x_n\equiv\ln\theta_n$ (angles measured in radians), $a\equiv=-1$ and
$b_n\equiv\ln[H_0 D_n/c]$.
For such tests ({\eg} \cite{Guerra00,Pen97}),
$\varepsilon_n$ includes scatter in the ``standard yardstick'' size.

For tests involving estimates $t_n$ of the age of the Universe
at redshifts $z_n$, we define $x_n\equiv\ln H_0 t_n$. Setting $a=1$ and $b_n=0$, this
gives \cite{Weinberg72}
\beq{AgeEq}
\fage =H_0\int_z^\infty{dz'\over(1+z')H(z')}.
\eeq

\Eq{xEq} implicitly assumes that the the measurement errors $\varepsilon_n$ for $x_n$ 
are additive. This is a popular approximation in many cases, notably for
 luminosity distance where the actual measurement is a magnitude proportional 
to $\ln d$, and tends to be accurate if the relative error 
(on flux, angular size, age, \etc) is small ($\ll 1$). In other cases, 
a more appropriate error model should be employed. 

Ideally, we would like to measure the function $\rhoeff(z)$ 
in ways that are sensitive only to the global spacetime geometry
and do not depend on $g(z,k)$.
The measurements that come closest to this ideal are arguably those 
involving the brightness, angular size and age of distant objects, in 
situations where the calibration of the ``standard candle'',
``standard ruler'' or ``standard clock'' depends mainly on microphysics.
However, great care must be taken when constraining competing theories of
gravity {\etc} (as opposed to more mundane dark energy), since 
slight time variations in microphysical constants could affect the
standard calibrators and masquerade as modifications to $\rhoeff(z)$.
For instance, time-variation in a non-gravitational ``constant'' (say the fine-structure
constant ``alpha'') could alter the intrinsic energetics and time-evolution
of a supernova explosion, making it a bad standard candle. If
such an effect were present but neglected, then
we would draw an incorrect conclusion about the luminosity distance
to the supernova and consequently also about the cosmic geometry.

Terrestrial constraints are are now so strong 
that time-variation of the fine structure constant $\alpha$ is unlikely to
cause such misinterpretations: variations in $\alpha$ are limited to 
around a part per million over cosmological 
timescales \cite{Damour96} (c.f. \cite {Webb99}).
A more serious concern in this context is 
time variations in the effective gravitational constant, 
expected in many scalar-tensor theories according to 
\eq{etaDefEq}.
Here the observational constraints are weaker, permitting variations of 
a few percent over 
cosmological time scales \cite{Dickey94,Williams96},
which could among other things cause distant supernovae
to be systematically brighter or fainter than nearby ones.

As \fig{rhoFig} indicates, the function $\rhoeff(z)$ is probed by
the growth of clustering as well, via the power
spectrum growth in the Ly$\alpha$ forest, weak lensing and galaxy surveys
as well as via the abundance evolution of galaxy clusters and lensing systems.
However, these probes all depend on the function $g(z,k)$ as well, \ie, 
on the next order of perturbation theory.
The Alcock-Paczynski test elegantly eliminates the dependence on $g(z,k)$
caused by number density evolution, but involves certain bias-related issues
\cite{AlcockPaczynski79,Nair99,Huterer00}.

Although the CMB power spectrum $C_\l$ depends on the growth $g$, 
the effect is ``vertical'' rather than ``horizontal''. In other words, 
as long as some detectable acoustic peak structure is visibly preserved,
it should be possible to extract the angular scale corresponding to
the horizon size at recombination fairly independently of $g$.
With additional knowledge of the dark matter density, 
this provides a clean measurement of $\flum$ at $z\sim 10^3$.

If GR is correct so that $g$ and the vertical structure of the
acoustic peak heights can be computed from first principles,
then the CMB is a sensitive probe of the high-redshift asymptote
in of the $\rhoeff$ curve in \fig{zdFig},
where the Universe becomes matter dominated.
Writing $\rho_m(z)=\rho_m^0 (1+z)^3$,
the MAP and Planck satellites
are expected to measure $\rho_m^0$ to accuracies of $10\%$ and $1\%$,
respectively. However, in contrast 
to, \eg, SN 1a, the CMB is mainly sensitive to 
the physical matter density $\rho_m$, not to the plotted quantity
$\rho_m/h^2$, since this is what matters for the acoustic
oscillations at $z>10^3$.
The ultimate accuracy with which we can measure $\rho_m^0/h^2$ (equivalently
$\Omega_m$) is therefore likely to be limited not by CMB issues
but by our ability to measure $h$, either directly or by combining 
various complementary measurements \cite{parameters2}.

\subsection{$\rhoeffz$ from existing SN 1a data}

Of the various low-redshift probes of $\rhoeff(z)$ that we 
discussed, type 1a supernovae are arguably the most sensitive
at the present time.
We will therefore compute constraints on $\rhoeff(z)$ from current 
SN 1a data and make forecasts for the accuracy attainable with
future measurements.

\begin{figure}[pbt]
\centerline{\epsfxsize=9.0cm\epsffile{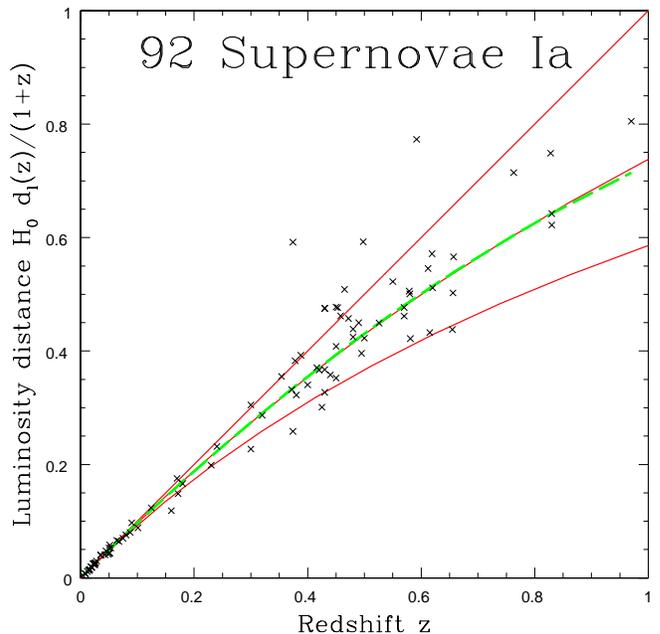}}
%\vskip-4.5cm
\caption{
Crosses show the luminosity distance for 92 SN 1a.
From top to bottom, solid curves correspond to models 
$(\Om,\Ol)=(0,1)$, $(0.38,0.62)$ and $(1,0)$, respectively.
The middle curve is almost indistinguishable from the best fit
quartic polynomial (dashed).
The density history $\rhoeff(z)$ is simply the squared inverse slope
of this curve.
Scatter increases with $z$ since the {\it relative} distance errors
are roughly constant.
}
\label{zdFig}
\end{figure}

\begin{figure}[pbt]
\vskip-1cm
\centerline{\epsfxsize=9.0cm\epsffile{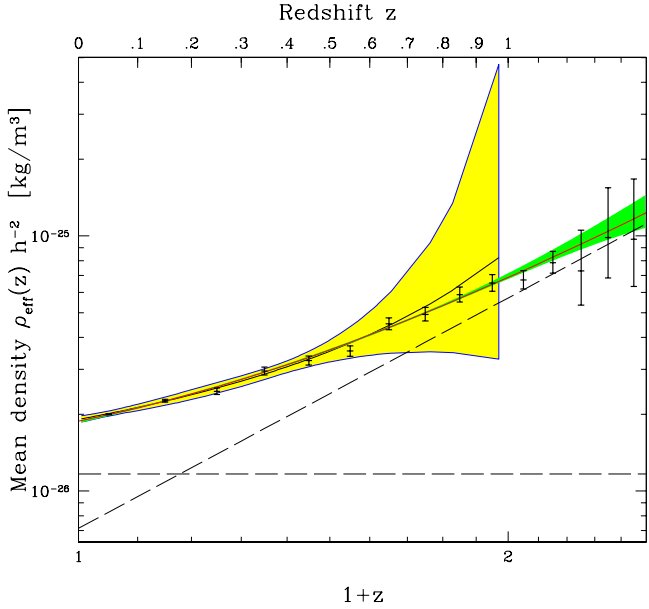}}
\vskip-0.5cm
\caption{
Zoom of \fig{rhoFig} showing constraints 
on $\rhoeff(z)$ from actual and simulated data, assuming a flat Universe.
Solid black curve shows best fit to the 92 SN1a,
corresponding to the polynomial fit shown in \fig{zdFig},
and yellow/light grey area shows the associated 
68\% confidence region. Green/dark grey area shows
the corresponding 68\% confidence region from our
SNAP simulation, for a fiducial model with
$\Om=0.38$, $\Ol=0.62$ (red/grey curve) whose 
two components are shown as dashed lines.
Error bars are for the non-parametric reconstruction
of \sec{SNAPsec},
spaced so that measurements of neighboring bands are uncorrelated,
and are identical to those shown in \fig{rhoFig}.
}
\label{zrhoFig}
\end{figure}

\Fig{zdFig} shows the measured luminosity distance
for 92 SN 1a, combining the high redshift sample of 42 
from \cite{Perlmutter98} with the 50 reported in 
\cite{Riess98} with the MLCS method.
Apart from systematic and calibration issues, 
our desired function $\rhoeff(z)$ is simply given by the
derivative of the curve around which they scatter:
$\rhoeff(z)=3H_0^2/8\pi\Geff\eta'(z)^2\propto h^2/\eta'(z)^2$.\footnote{
From here on, we will limit our discussion to the currently favored case
of flat space, \ie, $k=0$, where $\eta(z)=\flum(z)/(1+z)$
is directly measured for each supernova.
In the general case, 
$\eta(z) = k^{-1}\sin^{-1}[k\flum(z)/(1+z)]$ or 
$\eta(z) = k^{-1}\sinh^{-1}[k\flum(z)/(1+z)]$ depending on the sign of
the curvature, introducing an extra degeneracy with $k$ that has been 
exhaustively treated elsewhere 
\cite{WhiteComplementarity,complementarity,Huey99,parameters2}.
Since CMB experiments are measuring the 
acoustic peak locations to increasing precision, and these
are mainly sensitive to curvature (given by $\rhoeff(0)/h^2$),
degeneracies 
between different evolution scenarios for $\rhoeff(z)$ will be
more important than the curvature degeneracy in the post-MAP era.
}
Since the slope is seen to fall, the cosmic density 
clearly increases at high redshift. 
If standard GR is correct ($\rhoeff=\rho$), 
then this fact that $d\rhoeff/dz>0$
corresponds to the weak energy condition \cite{Wald84,Wang01} that the density 
measured by any observer is non-negative, and is equivalent to 
the pressure-to-density ratio constraint $p/\rho\ge -1$.
A more detailed 
reconstruction of $\rho(z)$ is shown in \fig{zrhoFig},
and was computed by fitting a homogeneous quartic polynomial
(dashed curve in \fig{zdFig})
to $\eta(z)$.
% with straightforward $\chi^2$-minimization,
% marginalizing over the offsets between the different sets of supernovae.
Using the notation of \eq{xEq}, the fit was done by minimizing
\beq{chi2Eq}
\chi^2 = \sum_{n=1}^n \left({x_n - a\ln d(z_n) - b_n\over\sigma_n}\right)^2,
\eeq
where $\sigma_n$ are the quoted 1 sigma magnitude errors \cite{Perlmutter98,Riess98}.
This gives the fitting parameters (polynomial coefficients) as linear combinations of 
the data $x_n$. 
The 1 sigma errors in \fig{zrhoFig} are obtained by propagating the errors on 
these fitting parameters.\footnote{\Eq{chi2Eq} assumes that
the magnitude errors on the different supernovae are uncorrelated. 
It is straightforward to generalize the procedure to the case of correlated errors,
where the $N\times N$ magnitude covariance matrix is non-diagonal given a physical model
for the nature of these correlations. A perfectly correlated term, corresponding to a common 
offset in all magnitudes, would have no effect on the results since it is only
the {\it derivative} that matters in \fig{zdFig}, but redshift-dependent offsets can of course
bias the results. Indeed, a deeper understanding of possible systematic errors of this type is
the main challenges for taking full advantage of the statistical power of 
future experiments like SNAP.
} 
The figure clearly illustrates the familiar fact that 
dark energy/modified gravity is required:
the slope of the $\rhoeff(z)$-curve is too shallow at low
redshift to be explainable in terms of ordinary
matter $(\rho\propto [1+z]^3$), spatial curvature
$(\rho\propto [1+z]^2$) or some combination of the two.

Since $\rhoeff(z)\propto h^2/\eta'(z)^2$, the SN 1a give us a direct measurement
of $\rhoeff(z)/h^2$, where $h$ is the Hubble parameter today 
in units of 100 km/s/Mpc.
Changing $h$ thus shifts $\rhoeff$ vertically
on a logarithmic plot like \fig{zdFig} without affecting 
its problematic slope.

\begin{figure}[pbt]
\vskip-1cm
\centerline{\epsfxsize=9.0cm\epsffile{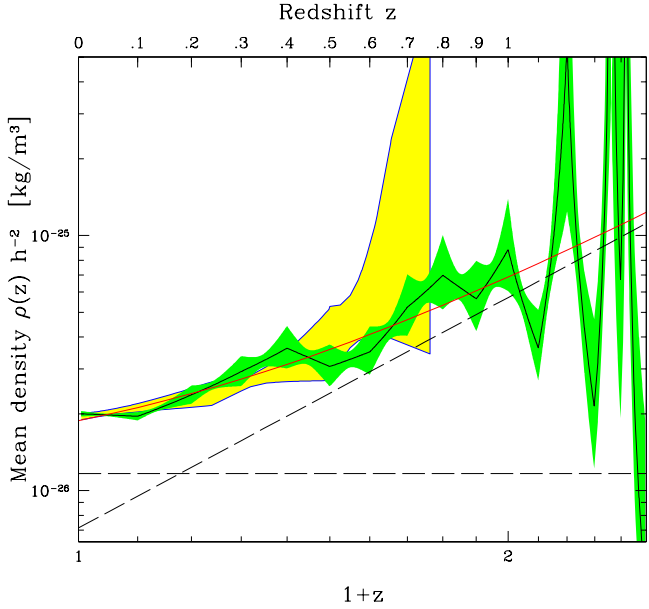}}
\vskip-0.5cm
\caption{
Same as previous figure,
%\protect\fig{zrhoFig}, 
but using a quadratic
spline instead of a polynomial to fit $\eta(z)$.
Bin widths were $\Delta z=0.2$ for the actual data
and $\Delta z=0.1$ for SNAP.
}
\label{zrhoFig1}
\end{figure}

\begin{figure}[pbt]
\vskip-1cm
\centerline{\epsfxsize=9.0cm\epsffile{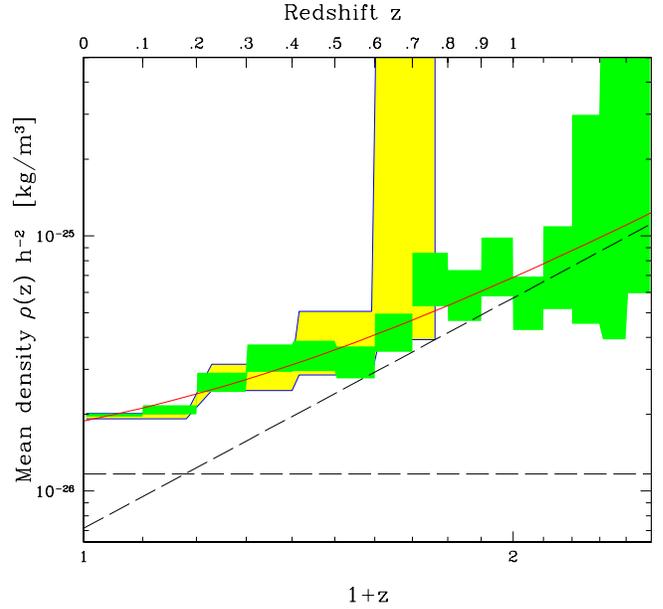}}
\vskip-0.5cm
\caption{
Same as previous figure but using linear
interpolation to fit $\eta(z)$, giving a piecewise constant density.
}
\label{zrhoFig2}
\end{figure}

To assess the sensitivity of our results to method details, we
show alternative analyses in figures~\ref{zrhoFig1} and~\ref{zrhoFig2}.
Here we replaced the polynominal fitting function
by piecewise quadratic and piecewise linear interpolating
functions for $\eta(z)$, respectively. 
For the quadratic case, we enforced that the function have
continuous derivative, thereby making $\rhoeff$ a continuous
function. The results are seen to be quite robust, in all cases
agreeing well with the ``concordance'' model and indicating a shallow
slope at low redshifts incompatible with just matter and curvature alone.
A very nice non-parametric SN 1a analysis was recently performed by
Wang \& Garnavich \cite{Wang01}, and it is interesting to compare the two 
studies since they were performed concurrently and independently.
The quantitative agreement is good taking into account that \cite{Wang01} 
treat the dark energy density $\rho(z)-\rho_m^0(1+z)^3$ rather 
than $\rhoeff(z)$ as the free function and marginalize 
over $\rho_m$ imposing the weak energy condition.
The SN 1a data thus speak loud and clear: the dark energy puzzle
is robust to theoretical assumptions and method details.

%%%%%%%%%%%%%%%%%%%%%%%%%%%%%%%%%%%%%%%%%%%%%%%%%%%%%%%%%
%%%%%%%%%%%%%%%%%%%%%%%%%%%%%%%%%%%%%%%%%%%%%%%%%%%%%%%%%

%%%%%%%%%%%%%%%%%%%%%%%%%%%%%%%%%%%%%%%%%%%%%%%%%%%%%%%%%
%%%%%%%%%%%%%%%%%%%%%%%%%%%%%%%%%%%%%%%%%%%%%%%%%%%%%%%%%

%%%%%%%%%%%%%%%%%%%%%%%%%%%%%%%%%%%%%%%%%%%%%%%%%%%%%%%%%

%%%%%%%%%%%%%%%%%%%%%%%%%%%%%%%%%%%%%%%%%%%%%%%%%%%%%%%%%

\subsection{With upcoming SN 1a data}
\label{SNAPsec}

The proposed SNAP satellite\footnote{More information on the SNAP mission is available at\\
$http://snap.lbl.gov$} would measure of order $N=2000$ SN 1a out to redshifts 
around 1.2 or higher \cite{Huterer00}.
There has been substantial debate in the literature about the accuracy with which the
equation of state can be reconstructed
\cite{Huterer99,Huterer00,Saini00,Kujat01,Maor02a,Maor02b,Weller02,Wasserman02,Spergel02,Peebles02,Friemann02,Linder02,Wang01}.
%Numerous forecasts have Forecasts have shown the exciting potential this has for probing the equation of 
%state of dark energy \cite{Huterer00,Wang01}. 
We will now compute how accurately SNAP can
measure $\rhoeff(z)$. It is straightforward to do this numerically
in a ``black box'' fashion, and figures~\ref{zrhoFig}-\ref{zrhoFig2}
show the result of doing this with simulations 
in the same way as we did it for the currently existing data.
However, because of the strong interest and substantial resources
focused on designing such future probes, it is also worthwhile to gain intuition on
how experimental specifications translate into model constraints.
We will therefore derive an analytic result which has the advantage of 
explicitly showing how the results scale when changing survey specifications,
the smoothing scale, \etc

\subsubsection{An analytic error formula}

Defining $f(z)\equiv H(0)/H(z)$, we have $\eta(z) = \int_0^z f(z')dz'$
and wish to know how accurately $f$ can be recovered from noisy
measurements of $\eta$. In other words, we simply want to compute the
derivative of a function measured with sparse sampling and noise.
The errors on $\rhoeff(z)\propto1/f(z)^2$ then follow trivially
from those on $f(z)$.

The accuracy with which $f(z)$ can be measured is 
readily computed with the Fisher information matrix formalism
\cite{Fisher35,karhunen} with infinitely many parameters
to be estimated (the value of $f$ at each $z$).
The amount of information about these parameters in the SN 1a data is 
given by \cite{complementarity}
\beq{Feq}
\F(z',z'') =  {a^2 N\over(\Delta m)^2}\int_0^\infty g(z) w_{z'}(z) w_{z''}(z) dz,
\eeq
where $a\equiv 5/\ln 10$,
$\Delta m$ is the rms SN 1a magnitude error, 
$N$ is the number of supernovae,
$g(z)$ is their redshift distribution (normalized to integrate to unity),
 and 
\beq{weq}
w_{z'}(z) \equiv {\partial\ln\eta(z)\over\partial f(z')} 
%= {\partial\over\eta(z)\partial f(z')}\int_0^z f(z'')dz''
= {\theta(z-z')\over\eta(z)}.
\eeq
since $\eta'(z)=f(z)$.
%using \eq{LumDistEq}.
Here $\theta$ denotes the Heaviside step function 
($\theta(z)=1$ for $z\ge 0$, vanishing otherwise).
\Eq{Feq} thus gives 
\beqa{Feq2}
\F(z,z')&=&\int\theta(z''-z)\theta(z''-z')h(z'')^{-1}dz''\\
        &=&\int_{{\rm max}(z,z')}^\infty h(z'')^{-1} dz'',\quad\hbox{where}\\
h(z)&\equiv& {(\Delta m)^2\over a^2 N}{\eta(z)^2\over g(z)}.
% = {a^2N\over(\Delta m)^2} \int_{{\rm max}(z',z'')}^\infty {g(z)\over\eta(z)^2} dz,
\eeqa
The covariance matrix $\C$ giving the best attainable error bars on 
our estimated parameters is the inverse of this infinite-dimensional matrix, 
$\C=\F^{-1}$. Fortunately, this inverse can be computed analytically, giving
\beq{Ceq}
\C(z,z') 
%\equiv\F^{-1}(z',z'')
= -h'(z)\delta'(z-z') - h(z)\delta''(z-z'),
% = - {(\Delta m)^2\over a^2 N} {\eta(z)^2\over g(z)} \delta''(z'-z''),
\eeq
where $\delta$ denotes the Dirac delta function.
This follows from the identity 
\beq{ProofEq}
\int\C(z,z'')\F(z'',z')dz'' = \delta(z-z'),
%&&-\int\int{h(z'')\over h(z)}\delta''(z-z''')\theta(z''-z''')\theta(z''-z')dz''dz'''
\eeq
which is proven by substituting equations\eqn{Feq2} and\eqn{Ceq},
integrating by parts twice and using that $\theta'(x)=\delta(x)$.
In practice we obviously want to smooth our estimated $f$-function ($\fest(z)$, say) 
to reduce noise. Using weighted averages 
$\fest_i\equiv \int\psi_i(z)\fest(z)dz$, the covariance matrix 
$\C_{ij}$
for these averages is
\beq{DiscreteCovEq}
%\C_{ij} = 
\int\C(z,z')\psi_i(z)\psi_j(z')dzdz' = \int h(z)\psi'_i(z)\psi'_j(z)dz,
\eeq
employing \eq{Ceq} and integrating by parts.
Once these errors on $f(z)$ have been computed, it is of course 
trivial to map them into errors on $\rhoeff(z)$. 
When the relative errors are small $(\Delta f\ll f)$,
we have simply $\Delta\rhoeff/\rhoeff\approx \Delta f/f$.

\subsubsection{Smoothing functions}

Using Gaussian smoothing functions 
\beq{GaussPsiEq}
\psi_i(z)\equiv 
{1\over (2\pi)^{1/2}\Delta z} e^{-{1\over 2}\left({z-z_i\over\Delta z}\right)^2},
\eeq
the error bars $\Delta f_i\equiv\C_{ii}^{1/2}$ are given by
\beq{fErrorEq}
\Delta f_i 
\approx {\ln 10\over 10\pi^{1/4}} 
{\Delta m\,\eta(z_i)\over (\Delta z)^{3/2}N^{1/2}g(z_i)^{1/2}}
\eeq
for the case where the smoothing scale $\Delta z$ is much smaller than
the scale on which $\eta$ and $g$ vary substantially.
In the same approximation,
the dimensionless correlation coefficients between the different
averages are 
\beq{rEq}
r_{ij}
%\equiv{\C_{ij}\over\sqrt{\C_{ii}\C_{jj}}}
\approx 
(2-n^2)e^{-n^2},\quad\hbox{where}\quad
n\equiv\left({z_i-z_j\over\Delta_z}\right)
\eeq
is the number of smoothing lengths by which the two redshifts 
are separated, so neighboring measurements are uncorrelated if
they are separated by $\sqrt{2}\Delta z$.
Another interesting case for our discussion below is 
that of ``triangular'' smoothing functions 
\beq{TriaPsiEq}
\psi_i(z) = 
\cases{
{\Delta z-|z-z_i|\over\Delta z} 	&if $|z-z_i|<\Delta z$,\crr
0					&otherwise,
} 
\eeq
which gives 
\beq{fErrorEq2}
\Delta f_i 
\approx {\sqrt{2}\ln 10\over 5} 
{\Delta m\,\eta(z_i)\over (\Delta z)^{3/2}N^{1/2}g(z_i)^{1/2}}
\eeq
in the same approximation. Here there are strong anticorrelations
$r=-2^{-1/2}\approx -0.7$ between neighboring bins, and all longer-range 
correlations vanish.
``Boxchar'' smoothing where we simply average in bins,
$\psi(z)=\theta(\Delta z/2-|z-z_i|)/\Delta z$, may seem like a natural
choice. This is a disaster, however, 
since the resulting $\delta$-functions in $\psi'_i$
blow up when squared. This problem is easy to understand physically. In a
toy example where we have many SN 1a equispaced in redshift
and estimate $\eta'(z)$ by simply differencing neighboring 
supernovae, averaging a segment of these derivative estimates
will reduce to simply differencing the first and the last, placing all
the statistical weight on merely two objects.

As we will return to below, our black-box numerical method can also 
understood in terms of smoothing functions of a particular form.

The exact result of \eq{DiscreteCovEq} and the approximation of
\eq{fErrorEq} illustrate a number of issues relevant to measuring 
$\rhoeff(z)$ in practice, as discusses in the following subsections.

\subsubsection{Redshift resolution}

Errors scale as $(\Delta z)^{-3/2}$ rather 
than $(\Delta z)^{-1/2}$ because of the derivative
nature of what we are measuring, which means that very large 
number of SN 1a are needed to probe the small-scale structure of $\rhoeff(z)$.
It also means that care must be taken in interpreting error forecasts to avoid 
coming away with a misleadingly pessimistic impression.
The superiority of SNAP over existing data is 
partly hidden in figures~\ref{zrhoFig1} and figures~\ref{zrhoFig2} since
$\Delta z$ is halved for SNAP, roughly tripling error bars.
The error bars from the numerical SNAP simulations in figures~\ref{zrhoFig1}
and figures~\ref{zrhoFig2} agree well with the analytic ones of 
\eq{TriaPsiEq} with $\Delta z=0.1$ as well as those of \cite{Wang01},
but are substantially larger than those from the polynomial fit shown in
\fig{zrhoFig}. This is because the quartic polynomial enforces more
smoothness, effectively increasing $\Delta z$.
The non-parametric error bars in \fig{zrhoFig} from \eq{fErrorEq}
using Gaussian smoothing
are a factor of $2^{3/2}\pi^{1/4}\sim 4$ smaller than
those from \eq{fErrorEq2} using triangle smoothing,
which is related to the fact that the triangle smoothing functions
have standard deviations $\sqrt{6}$ times narrower than the Gaussian ones.
% when rescaling triangle and gaussian smoothing functions to have the
% same standard deviation, the resulting error bars $\Delta\fest$
% differ by merely 4%.

Which error bars are most appropriate depends on what the data are to be used for.
Constraining models where $\rhoeff$ may change abruptly with $z$ requires
high redshift resolution and associated large error bars like in
fugures~\ref{zrhoFig1} and~\ref{zrhoFig2}.
Most published models, however, predict rather smooth 
functions $\rhoeff(z)$, so the small error bars from 
a low-order polynomial fit like in \fig{zrhoFig} 
give the most accurate representation
of how well SNAP could distinguish them from one another.

\subsubsection{High-$z$ sensitivity}
  
Since errors scale as $\eta(z)$, one does worse at large $z$.
However, since $\eta(z)$ flattens out around $z\sim 1$ and 
asymptotes to a constant as $z\to\infty$, this is not a fundamental 
limitation to probing very large redshifts --- the challenge is simply
to find suitable high-$z$ standard candles or other reference objects to prevent 
$g(z)$ from going to zero in the denominator.

\subsubsection{Optimal estimation}

\Eq{DiscreteCovEq} also indicates what the optimal estimator
is for non-parametric measurement of $\rhoeff(z)$. 
It is easy to verify that when the smoothing scale $\Delta z$ is 
much smaller than the
scale $h/|h'|$ on which $h$ varies substantially, 
weighting the actual data with the function $\psi_i'(z) h(z)$ and normalizing 
appropriately will recover the minimal Fisher matrix error bars.
In practice, $g(z)$ is of course not a smooth function but a 
sum of $\delta$-functions at the observed SN 1a redshifts,
so an additional correction for the 
non-even spacing between adjacent SN 1a is necessary.
% And I haven't yet figured out how to do this...

\begin{figure}[pbt]
%\vskip-2cm
\centerline{\epsfxsize=9.0cm\epsffile{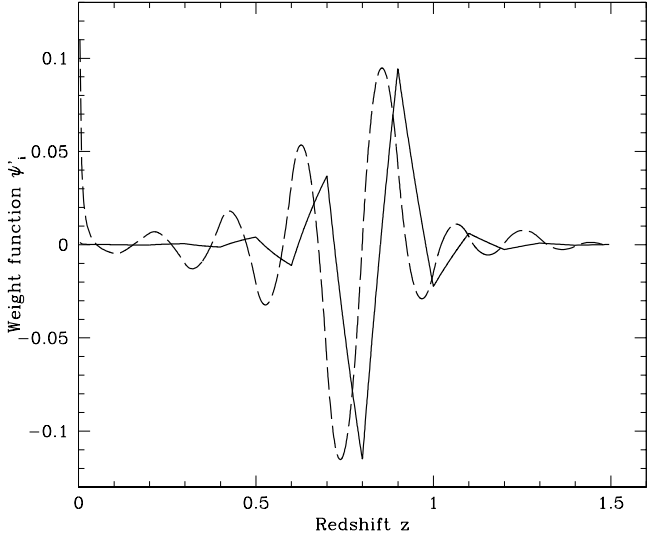}}
%\vskip-4.5cm
\caption{
Sample weight functions $\psi'(z)$ extracted from the $\chi^2$-minimization
are plotted for the quadratic spline (dashed) and linear interpolation (solid),
both probing the derivative $\eta'(z)$ at $z\sim 0.8$.
Excessive ringing drives up the error bars and the fact that the functions
are so wide complicates the interpretation.
}
\label{windowFig}
\end{figure}

The black box of $\chi^2$-minimization can also be understood 
in terms of smoothing functions and \eq{DiscreteCovEq}.
Since $\chi^2$ depends quadratically on the free parameters, 
the best fit parameters and hence also the recovered function $\fest(z)$ depends 
linearly on the observed $\eta$-values for each supernova, once the redshifts are given.
In other words, we can write 
\beq{NumericalPsiEq}
\fest_i = \sum_{j=1}^N \psi'_i(z_j) \eta_j
\eeq
for some function $\psi_i$, where $\eta_j$ is the observed 
$\eta$-value for the $j^{th}$ supernova.
In an attempt to demystify the numerical results somewhat, 
we have plotted these weight functions $\psi'_i$ in
\fig{windowFig} for estimating $f(z)$ with SNAP at $z\sim 0.8$.
We see that for both the quadratic spline and the linear interpolation
cases, there is substantial ringing. Moreover, the
measurement of $\eta'(0.8)$ is seen to involve supernovae over a
very broad range of redshifts, not merely near $z=0.8$.
In contrast, the triangle smoothing function would be more local,
giving $\psi'(z)=-1$ for $z=0.7-0.8$, $\psi'(z)=1$ for $z=0.8-0.9$ 
and zero elsewhere, simply estimating the derivative $\eta'(0.8)$
by subtracting measurements to the left from ones to the right.
The Gaussian smoothing function behaves similarly, merely with
less abrupt behavior. Just as sophisticated data analysis techniques
developed by numerous groups have enhanced the science return
of recent CMB missions, there may well be room for methodological
improvement here. For instance, what is a good parametrization of 
$f$ that leads to compact, easy-to-interpret weight functions?
% giving small error bars saturating the Fisher matrix bound.

%%%%%%%%%%%%%%%%%%%%%%%%%%%%%%%%%%%%%%%%%%%%%%%%%%%%%%%%%

%%%%%%%%%%%%%%%%%%%%%%%%%%%%%%%%%%%%%%%%%%%%%%%%%%%%%%%%%

\section{DISCUSSION}
\label{DiscussionSec}

We have argued that the effective density evolution 
$\rhoeff(z)$ and the linear growth factor $g(z,k)$ 
constitute the natural meeting point for theory
to confront observation in terms of linear cosmology,
since they can be measured with very few assumptions
about unseen matter and the underlying theory of gravity.
This generalizes studies of the equation of state $w$.
 
We analyzed current SN 1a data in this framework, and showed that
the statistical constraints on $\rhoeff(z)$ are already 
fairly strong for $z\simlt 0.5$.
This alternative approach confirms the results of 
previous SN 1a analyses \cite{Riess98,Perlmutter98,WhiteComplementarity,Wang01}, and
provides a clean and model-independent case for either 
dark energy or modified gravity, since $\rhoeff(z)$ has 
the wrong logarithmic slope at low redshift to be explained
by either matter (slope $-3$) or spatial curvature (slope $-2$).
 
We estimated the accuracy with which the proposed SNAP satellite should
be able to measure $\rhoeff(z)$, both with a numerical simulation 
involving polynomial fitting and by deriving a non-parametric 
analytic formula, and found that precision density measurements 
should be attainable in about a dozen independent redshift bins
out to $z\sim 1.2$. Since errors grow as $(\Delta z)^{-3/2}$ as the
bin width $\Delta z$ is reduced, the large numbers of 
SN 1a of SNAP are absolutely necessary to resolve the small-scale
structure of $\rhoeff(z)$. The reason that $w(z)$ is so much harder to constrain accurately
is that it involves a second derivative of the data, causing error bars to blow up as 
 $(\Delta z)^{-5/2}$.

%%%%%%%%%%%%%%%%%%%%%%%%%%%%%%%%%%%%%%%%%%%%%%%%%%%%
\begin{figure}[phbt]
\centerline{{\vbox{\epsfxsize=8.6cm\epsfbox{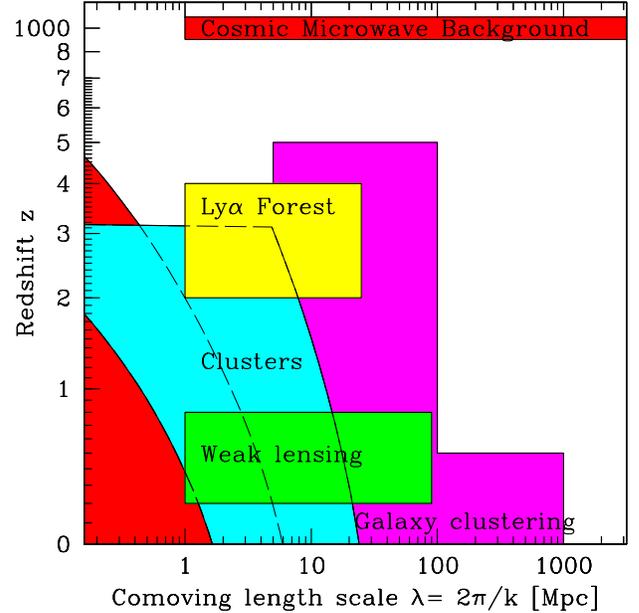}}}}
\smallskip\smallskip
\caption{
Shaded regions show crude estimates of the ranges of scale and redshift over which
various observations are likely to probe the growth factor 
$g(k,z)$ over the next few years.
The lower left region, delimited by the dashed line, is the non-linear
regime where rms density fluctuations exceed unity
for the ``concordance'' model from 
\protect\cite{concordance}.
% making theoretical predictions less reliable. 
}
\label{zkFig}
\end{figure}
%%%%%%%%%%%%%%%%%%%%%%%%%%%%%%%%%%%%%%%%%%%%%%%%%%%%

We focused on the 0th order function $\rhoeff(z)$, with the 1st order function
$g(z,k)$ studied in detail in a companion paper \cite{pwindows}.
To place our discussion in its observational context, 
\fig{zkFig} shows the rough ranges of scale and redshift over which
various observations are likely to probe $g(k,z)$ over the next few years.
Regardless of what detailed assumptions are made, it is clear from
figures~\ref{rhoFig} and~\ref{zkFig} that 
upcoming measurements of the CMB, the Ly$\alpha$ Forest, 
weak lensing, cluster abundance and galaxy clustering have the potential
to probe both functions back to high redshift over several orders of 
magnitude in scale, allowing numerous consistency checks between 
different types of measurements.
Measuring $g(z)$ for pressureless matter could provide an independent
determination of $H(z)$ \cite{Starobinski98}. 

It may be possible to extract further constraints on 
dark energy/modified gravity 
from going to higher order in 
perturbation theory or studying the nonlinear regime.
Such nonlinear issues were indeed the motivation for some recent suggestions
for dark matter and modified gravity \cite{Milgrom83,Milgrom98,Mannheim00}. 
Although many proposed quintessence fields do not cluster on 
small scales, early work in this direction 
\cite{Esposito00,Gaztanaga00} 
has discussed how temporal and perhaps spatial variations in 
Newton's constant may be an observable signature.

In conclusion, the evidence for dark energy/modified gravity 
has emerged as one of the key puzzles of modern cosmology.
Fortunately, there are at least half a dozen observational approaches
that can directly probe the time-evolution of the effective mean 
density and clustering, so by going after this problem 
with the full arsenal, there is real hope that the cosmology 
community can resolve it over the next decade.

\bigskip
The author wishes to thank Arthur Kosowsky and Jim Peebles for 
provocative questions triggering this work, 
and Bernd Br\"ugman, Ang\'elica de Oliveira-Costa, 
Daniel Eisenstein, Gilles Esposito-Farese, Wayne Hu,
Dragan Huterer, Arthur Kosowsky, 
David Polarski, Alexei Starobinski,  
Paul Steinhardt, Mike Turner and Robert Wald 
for helpful comments.

Support for this work was provided by
NSF grant AST00-71213, 
NASA grant NAG5-9194 and
the University of Pennsylvania Research Foundation.

%%%%%%%%%%%%%%%%%%%%%% REFERENCES: %%%%%%%%%%%%%%%%%%%%%%%%%

%\clearpage
%\end{multicols}

%\vskip-1.0cm


\begin{references}   % 

%\bigskip


%%%%%%%%%%%%%%%%%%%%%%%%%%%%%%%%%%%%%%%%%%%%%%%%%%%%%%%%%%%%%%%%%%%%%
% RECENT PARAMETER ESTIMATION STUFF:
%%%%%%%%%%%%%%%%%%%%%%%%%%%%%%%%%%%%%%%%%%%%%%%%%%%%%%%%%%%%%%%%%%%%%

\bibitem{Lange00}
\rf\nnn Lange A E {\etal};2001;Phys. Rev. D;63;042001
%astro-ph/0005004

\bibitem{boompa}
\rf\nn Tegmark M\dualand\nn Zaldarriaga M;2000;Phys. Rev. Lett.;85;2240    
% astro-ph/0004393

\bibitem{Bambi00}
\rf\nn Balbi A {\etal};2000;ApJL;545;L1
% astro-ph/0005124

\bibitem{Bridle00}
\rf\nnn Bridle S L {\etal};2001;MNRAS;321;333
% astro-ph/0006170

\bibitem{observables}
\rf\nn Hu W, \nn Fukugita M, 
\nn Zaldarriaga M\multiand\nn Tegmark M;2001;ApJ;549;669
% astro-ph/0006436
% Title: CMB Observables and Their Cosmological Implications
% Authors: Wayne Hu (IAS, Princeton), Masataka Fukugita, Matias Zaldarriaga (IAS, Princeton), Max Tegmark (U. Penn)
% Comments: 12pages, 11figs. Submitted to ApJ

\bibitem{Jaffe00}
\rf\nn Jaffe A {\etal};2000;Phys. Rev. Lett.;86;3475
% astro-ph/0007333
% Title: Cosmology from Maxima-1, BOOMERaNG and COBE/DMR CMB Observations
% Authors: A.H. Jaffe, P.A.R. Ade, A. Balbi, J.J Bock, J.R. Bond, J. Borrill, A. Boscaleri, K. Coble, B.P. Crill, P. de Bernardis, P.
% Farese, P.G. Ferreira, K. Ganga, M. Giacometti, S. Hanany, E. Hivon, V.V. Hristov, A. Iacoangeli, A.E. Lange, A.T. Lee, L. Martinis,
% S. Masi, P.D. Mauskopf, A. Melchiorri, T. Montroy, C.B. Netterfield, S. Oh, E. Pascale, F. Piacentini, D. Pogosyan, S. Prunet, B.
% Rabii, S. Rao, P.L. Richards, G. Romeo, J.E. Ruhl, F. Scaramuzzi, D. Sforna, G.F. Smoot, R. Stompor, C.D. Winant, J.H.P. Wu
% Comments: 5 Pages, 2 Figures, Submitted to PRL, Revtex

\bibitem{Kinney01}
\rf\nnn Kinney W H, 
\nn Melchiorri A\multiand\nn Riotto A;2001;Phys. Rev. D;63;23505
% astro-ph/0007375
% Title: New Constraints on inflation from the Cosmic Microwave Background
% Authors: William H. Kinney, Alessandro Melchiorri, Antonio Riotto
% Comments: 16 pages, 4 figures, some typos corrected, references added
% Journal-ref: Phys.Rev. D63 (2001) 023505

\bibitem{Durrer00}
% astro-ph/0009057
\rf\nn Durrer R\dualand\nn Novosyadlyj B;2001;MNRAS;324;560
% Title: Cosmological parameters from complementary observations of the Universe
% Authors: R. Durrer (Geneva University), B. Novosyadlyj (L'viv University)
% Comments: 12 pages, 3 figures. Submitted to MNRAS; minor corrections in the text (misprints, some
% values) and references
        
\bibitem{concordance}
\rf\nn Tegmark M, 
\nn Zaldarriaga M\multiand\nnnn Hamilton A J S;2001;Phys. Rev. D;63;43007
% astro-ph/0008167
% Title: Towards a refined cosmic concordance model: joint 11-parameter constraints from CMB and large-scale
% structure
% Authors: Max Tegmark, Matias Zaldarriaga, Andrew J. S. Hamilton
% Journal-ref: Phys.Rev. D63 (2001) 043007

\bibitem{consistent}
\rf\nn Wang X, \nn Tegmark M\multiand\nn Zaldarriaga M;2002;Phys. Rev. D;65;123001
%astro-ph/0105091

\bibitem{Efstathiou02}
\rf\nn Efstathiou G {\etal};2002;MNRAS;330L;29
% astro-ph/0109152
%    Title: Evidence for a non-zero Lambda and a low matter density from a combined
% analysis of the 2dF Galaxy Redshift Survey and Cosmic Microwave Background Anisotropies
% Authors: George Efstathiou, Stephen Moody, John A. Peacock, Will J. Percival,
% CarltonBaugh, Joss Bland-Hawthorn, Terry Bridges, Russell Cannon, Shaun Cole,
% Matthew Colless, Chris Collins, Warrick Couch, Gavin Dalton, Roberto De Propis,
% Simon P. Driver, Richard S. Ellis, Carlos S. Frenk, Karl Glazebrook, Carole Jackson
,


%%%%%%%%%%%%%%%%%%%%%%%%%%%%%%%%%%%%%%%%%%%%%%%%%%%%%%%%%%%%%%%%%%%%%

\bibitem{Scholberg99}
\rfprep\nn Scholberg K {\etal};1999;hep-ex/9905016
% Title: Atmospheric Neutrinos at Super-Kamiokande
% Authors: K. Scholberg, for the Super-Kamiokande Collaboration
% Comments: Talk presented at 8th International Workshop on Neutrino Telescopes, Venice, February 23-26 1999          
% THESE GUYS NEVER SEEM TO PUBLISH ANYTHING EXCEPT IN PROCEEDINGS!


%%%%%%%%%%%%%%%%%%%%%%%%%%%%%%%%%%%%%%%%%%%%%%%%%%%%%%%%%%%%%%%%%%%%%
% SOME SELF-INTERACTING DARK MATTER REFS:
%%%%%%%%%%%%%%%%%%%%%%%%%%%%%%%%%%%%%%%%%%%%%%%%%%%%%%%%%%%%%%%%%%%%%

\bibitem{Carlson92}
\rf\nnn Carlson M E, \nnn Machacek M E\multiand\nnn Hall L J;1992;ApJ;398;43
% Self-interacting dark matter
% Proposed self-interacting warm dark matter

\bibitem{deLaix95}
\rf\nnn {de Laix} A A, \nnn Scherrer R J\multiand\nnn Schaefer R K;1995;ApJ;452;495
% de Laix, Andrew A.;
% Scherrer, Robert J.;
% Schaefer, Robert K.
% Constraints on Self-interacting Dark Matter

%\bibitem{Atrio97}
%\rf\nn {Atrio-Barandela} F\dualand\nn Davidson S;1997;Phys. Rev. D;55;5886
% Atrio-Barandela, Fernando;
% Davidson, Sacha
% Interacting hot dark matter
						
\bibitem{Spergel00}
\rf\nnn Spergel D N\dualand\nnn Steinhardt P J;2000;Phys. Rev. Lett.;84;3760
% Observational evidence for self-interacting cold dark matter.

\bibitem{Hogan00}
\rf\nnn Hogan C J\dualand\nnn Dalcanton J J;2000;Phys. Rev. D;62;063511
% astro-ph/0002330
%Title: New Dark Matter Physics: Clues from Halo Structure
%Authors: Craig J. Hogan, Julianne J. Dalcanton (U. Washington)
% Constraints on WDM, SIDM

\bibitem{Hannestad00}
\rf\nn Hannestad S\dualand\nnn Scherrer R J;2000;Phys. Rev. D;62;043522
% astro-ph/0003046
% Title: Self-interacting Warm Dark Matter
% Authors: Steen Hannestad, Robert J. Scherrer

\bibitem{Burkert00}
\rf\nn Burkert A;2000;ApJL;534;L143 
% Burkert, Andreas
% The Structure and Evolution of Weakly Self-interacting Cold Dark Matter Halos

\bibitem{Firmani00}
\rf\nn Firmani C, \nn {D'Onghia} E, \nn {Avila-Reese} V, 
\nn Chincarini G\multiand\nn Hern\'andez X;2000;MNRAS;315;L29
% Firmani, C.; D'Onghia, E.;
% Avila-Reese, V.; Chincarini, G.;
% Hernández, X.
% Evidence of self-interacting cold dark matter from galactic to
% galaxy cluster scales

\bibitem{Mo00}
\rf\nnn Mo H J\dualand\nn Mao S;2000;MNRAS;318;163
% Mo, H. J.; Mao, Shude
%The Tully-Fisher relation and its implications for the halo density
%profile and self-interacting dark matter
						
\bibitem{Kochanek00}
\rf\nnn Kochanek C S\dualand\nn White M;2000;ApJ;543;514
% Kochanek, C. S.; White, Martin
% A Quantitative Study of Interacting Dark Matter in Halos
				
\bibitem{Yoshida00}
\rf\nn Yoshida N, \nn Springel V\multiand\nnnn White S D M;2000;ApJL;544;L87
% Yoshida, Naoki;
% Springel, Volker;
% White, Simon D. M.;
%Weakly Self-interacting Dark Matter and the Structure of Dark Halos
 
%%%%%%%%%%%%%%%%%%%%%%%%%%%%%%%%%%%%%%%%%%%%%%%%%%%%%%%%%%%%%%%%%%%%%%%%%%%%%%%%%%%%%%%%%%%%% 
 
\bibitem{Kaplinghat00}
\rf\nn Kaplinghat M, \nn Knox L\multiand\nnn Turner M S;2000;Phys. Rev. Lett.;85;3335
%astro-ph/0005210
%Title: Annihilating Cold Dark Matter
%Authors: M. Kaplinghat, L. Knox, M.S. Turner
%Comments: Matches version accepted for publication in Physical Review Letters. Included the
%effect of adiabatic expansion. Some changes in the abstract. Conclusions unchanged
%Journal-ref: Phys.Rev.Lett. 85 (2000) 3335
%SLAC-comments: Published in Phys.Rev.Lett.85:3335,2000 
        
%%%%%%%%%%%%%%%%%%%%%%%%%%%%%%%%%%%%%%%%%%%%%%%%%%%%%%%%%%%%%%%%%%%%%
% SOME WARM DARK MATTER REFS:
%%%%%%%%%%%%%%%%%%%%%%%%%%%%%%%%%%%%%%%%%%%%%%%%%%%%%%%%%%%%%%%%%%%%%

\bibitem{Bonometto85}
\rf\nnn Bonometto S A\dualand\nn Valdarnini R;1985;ApJL;299;L71
%Microwave background anisotropies, large-scale peculiar velocity
%fields, and clustering evolution in a warm-hot dark matter
%cosmological model

\bibitem{Schaeffer88}
\rf\nn Schaeffer R\dualand\nn Silk J;1988;ApJ;332;1
%Schaeffer, Richard;
%Silk, Joseph
%Cold, warm, or hot dark matter - Biased galaxy formation and pancakes

\bibitem{Colombi96}
\rf\nn Colombi S, \nn Dodelson S\multiand\nnn Widrow L M;1996;ApJ;458;1
%Colombi, Stephane;
%Dodelson, Scott;
%Widrow, Lawrence M.
%Large-Scale Structure Tests of Warm Dark Matter

\bibitem{Colin00}
\rf\nn {Col\'{\i}n} P, \nn {Avila-Reese} V\multiand\nn Valenzuela O;2000;ApJ;542;622
%Colín, Pedro;
%Avila-Reese, Vladimir;
%Valenzuela, Octavio
%Substructure and Halo Density Profiles in a Warm Dark Matter Cosmology

\bibitem{Narayanan00}
\rf\nnn Narayanan V K, \nnn Spergel D N, \nn {Dav\'e} R\multiand\nnn Ma C P;2000;ApJL;543;L103
%Narayanan, Vijay K.;
%Spergel, David N.;
%Davé, Romeel; Ma, Chung-Pei
%Constraints on the Mass of Warm Dark Matter Particles and the
%Shape of the Linear Power Spectrum from the LY&alpha; Forest
	
\bibitem{Bode00}
\rf\nn Bode P, \nnn Ostriker J P\multiand\nn Turok N;2001;ApJ;556;93
% astro-ph/0010389
% Halo Formation in Warm Dark Matter Models

%%%%%%%%%%%%%%%%%%%%%%%%%%%%%%%%%%%%%%%%%%%%%%%%%%%%%%%%%%%%%%%%%%%%%

\bibitem{Hu00}
\rf\nn Hu W, \nn Barkana R\multiand Gruzinov A;2000;Phys. Rev. Lett.;85;1158
% astro-ph/0003365
% Title: Cold and Fuzzy Dark Matter
% Authors: Wayne Hu, Rennan Barkana, Andrei Gruzinov (IAS, Princeton)
% Comments: 4 pages, 2 figures, submitted to PRL

\bibitem{Peebles00a}
\rf\nnnn Peebles P J E;2000;ApJL;534;127
% Fluid Dark Matter


%%%%%%%%%%%%%%%%%%%%%%%%%%%%%%%%%%%%%%%%%%%%%%%%%%%%%%%%%%%%%%%%%%%%%
% QUINTESSENCE REFS
%%%%%%%%%%%%%%%%%%%%%%%%%%%%%%%%%%%%%%%%%%%%%%%%%%%%%%%%%%%%%%%%%%%%%

\bibitem{Zlatev}
\rf\nn Zlatev I, \nn Wang L\multiand\nnn Steinhardt P J;1999;Phys. Rev. Lett.;82;896
% Zlatev, Ivaylo; Wang, Limin; Steinhardt, Paul J.
% Quintessence, Cosmic Coincidence, and the Cosmological Constant

\bibitem{SahniStarobinski00}
\rf\nn Sahni A\dualand\nnn Starobinsky A A;2000;Int. J. Mod. Phys. D;4;373
% International Journal of Modern Physics D, Volume 9, Issue 04, pp. 373-443 (2000).
% Useful review of phenomenological dark energy models. Table 1 is cute.
% In filing cabinet under "gravity".

%%%%%%%%%%%%%%%%%%%%%%%%%%%%%%%%%%%%%%%%%%%%%%%%%%%%%%%%%%%%%%%%%%%%%
% PEOPLE WHO WORRY:
%%%%%%%%%%%%%%%%%%%%%%%%%%%%%%%%%%%%%%%%%%%%%%%%%%%%%%%%%%%%%%%%%%%%%

\bibitem{Peebles99}
\rfprep\nnnn Peebles  P J E;1999;astro-ph/9910234
%Title: Summary: Comments on the State of our Subject
%Comments: 8 pages; to be published in Clustering at High Redshift, Marseilles, June 1999; eds. A.
%Mazure and O. Le Fevre

\bibitem{Peebles00}
\rfprep\nnnn Peebles  P J E;2000;astro-ph/0011252 
% Comments: 8 pages; in IAU Symposium 201, New Cosmological Data and the Values of the
% Fundamental Parameters, Manchester, August 2000, eds. A. N. Lasenby, A. W. Jones and A. Wilkinson

\bibitem{Sellwood00}
\rfprep\nnn Sellwood J A\dualand\nn Kosowsky A;2000;astro-ph/0009074
% Title: Does Dark Matter Exist?
% Comments: To appear in "Gas & Galaxy Evolution" eds Hibbard, Rupen & van Gorkom, 8 pages,

\bibitem{Disney00}
\rf\nnn Disney M J;2000;Gen. Rev. \& Grav.;32;1125
% astro-ph/0009020 
% General Relativity and Gravitation, Volume 32, Issue 6, p.1125-1134
% The Case Against Cosmology
%Journal-ref: GReGr (2000) 32 1125
%It is argued that some of the recent claims for cosmology are grossly overblown. Cosmology
%rests on a very small database: it suffers from many fundamental difficulties as a science (if it
%is a science at all) whilst observations of distant phenomena are difficult to make and harder
%to interpret. It is suggested that cosmological inferences should be tentatively made and
%sceptically received. 

\bibitem{McGaugh00}
\rf\nnn McGaugh S S;2000;ApJL;541;L33
% McGaugh, Stacy S.

%%%%%%%%%%%%%%%%%%%%%%%%%%%%%%%%%%%%%%%%%%%%%%%%%%%%%%%%%%%%%%%%%%%%%
% RECENT MODIFIED GRAVITY SUGGESTIONS:
%%%%%%%%%%%%%%%%%%%%%%%%%%%%%%%%%%%%%%%%%%%%%%%%%%%%%%%%%%%%%%%%%%%%%

\bibitem{Milgrom83}
\rf\nn Milgrom M;1983;ApJ;270;365
%A modification of the Newtonian dynamics as a possible
%alternative to the hidden mass hypothesis

\bibitem{Milgrom98}
\rfprep\nn Milgrom M;1998;astro-ph/9810302 
%The Modified Dynamics--A Status Review
%Authors: Mordehai Milgrom (Weizmann Institute)
%Comments: 15 pages, latex, 3 embedded figures. To appear in the proceedings of the Second
%International Workshop on Dark Matter (DARK98), eds. H.V. Klapdor-Kleingrothaus and L. Baudis

\bibitem{Damour99}
\rfprep\nn Damour T;1999;gr-qc/9904057
%From: damour@ihes.fr
%Date: Thu, 22 Apr 1999 13:34:07 GMT   (15kb)
%Experimental Tests of Relativistic Gravity
%Authors: Thibault Damour
%Comments: 10 pages, latex, uses espcrc2.sty, invited talk at the 1998 Texas Symposium (to appear in
%Nucl. Phys. B (Proceedings Supplement))
%Report-no: IHES/P/99/31
%Journal-ref: Nucl.Phys.Proc.Suppl. 80 (2000) 41-50
%The confrontation between Einstein's gravitation theory and experimental results, notably
%binary pulsar data, is summarized and its significance discussed. Experiment and theory agree
%at the 10^{-3} level or better. All the basic structures of Einstein's theory (coupling of
%gravity to matter; propagation and self-interaction of the gravitational field, including in
%strong-field conditions) have been verified. However, the theoretical possibility that scalar
%couplings be naturally driven toward zero by the cosmological expansion suggests that the
%present agreement between Einstein's theory and experiment might be compatible with the
%existence of a long-range scalar contribution to gravity (such as the dilaton field, or a moduli
%field, of string theory). This provides a new theoretical paradigm, and new motivations for
%improving the experimental tests of gravity. 

\bibitem{Mannheim00}
\rf\nn Mannheim P;2000;Found. Phys.;30;709
%gr-qc/0001011 
%From: Philip Mannheim <mannheim@phys.uconn.edu>
%Date: Thu, 6 Jan 2000 13:20:06 GMT   (36kb)
%Attractive and Repulsive Gravity
%Author: Philip D. Mannheim (University of Connecticut)
%Comments: RevTeX, 31 pages. Prepared for Foundations of Physics Festschrift in honor of Kurt Haller
%Journal-ref: Found.Phys. 30 (2000) 709-746
%We discuss the circumstances under which gravity might be repulsive rather than attractive.
%In particular we show why our standard solar system distance scale gravitational intuition
%need not be a reliable guide to the behavior of gravitational phenomena on altogether larger
%distance scales such as cosmological, and argue that in fact gravity actually gets to act
%repulsively on such distance scales. With such repulsion a variety of current cosmological
%problems (the flatness, horizon, dark matter, universe age, cosmic acceleration and
%cosmological constant problems) are then all naturally resolved. 
    
\bibitem{Boisseau00}
\rf\nn Boisseau B, \nn {Esposito-Far\`ese} G, 
\nn Polarski D\multiand\nnn Starobinsky A A;2000;Phys. Rev. Lett.;85;2236
% Cosmic acceleration and scalar-tensor gravity

\bibitem{Esposito00}
\rf\nn {Esposito-Far\`ese} G\dualand\nn Polarski D;2001;Phys. Rev. D;63;063504
% gr-qc/0009034 
%Scalar-tensor gravity in an accelerating universe
%Authors: G. Esposito-Farese, D. Polarski
%Comments: 37 pages, LaTeX 2.09, REVTeX 3.0, uses epsf.tex to include 6 postscript figures

\bibitem{Gaztanaga00}
\rf\nn {Gazta\~naga} E\dualand\nn Lobo A;2001;ApJ;548;47
% 47-59
% astro-ph/0003129
% Title: Non-Linear gravitational growth of large scale structures inside and outside standard
% Cosmology
% Authors: Enrique Gaztanaga, Alberto Lobo
% Comments: 15 pages, 9 figures, LaTeX, accepted in the Astrophysical Journal. Replaced with
% minor changes, a new figure with Cluster Abundance predictions added

% Two in response to Bernardeau's "cite me" email:

\bibitem{Binetruy00}
\rf\nn Binetruy P\dualand\nn Silk J;2001;Phys. Rev. Lett.;87;031102
% astro-ph/0007452
% Authors: Pierre Binetruy, Joseph Silk

\bibitem{Uzan00}
\rf\nn Uzan J P\dualand\nn Bernardeu F;2001;Phys.Rev. D;64;083004
% hep-ph/0012011
% Lensing at cosmological scales: a test of higher dimensional gravity
% Authors: Jean-Philippe Uzan (LPT, Orsay), Francis Bernardeau (SPhT, CEA Saclay)

\bibitem{Hwang02}
\rf\nn Hwang J\dualand\nn Noh H;2002;Phys. Rev. D;65;023512
% astro-ph/0102005
% Gauge-ready formulation of the cosmological kinetic theory in generalized gravity theories

\bibitem{Behnke02}
\rf\nn Behnke D, \nn Blaschke D, 
\nnn Pervushin V N\multiand\nn Proskurin D;2002; Phys. Lett. B;530;20
% gr-qc/0102039
% Description of Supernova Data in Conformal Cosmology without Cosmological Constant


%%%%%%%%%%%%%%%%%%%%%%%%%%%%%%%%%%%%%%%%%%%%%%%%%%%%%%%%%%%%%%%%%%%%%
%% PPN etc:
%%%%%%%%%%%%%%%%%%%%%%%%%%%%%%%%%%%%%%%%%%%%%%%%%%%%%%%%%%%%%%%%%%%%%

\bibitem{WillBook}
\rfbook\nnn Will C M;1993;Theory and Experiment in Gravitational 
Physics;Cambridge University Press;Cambridge

\bibitem{Will98}
\rfprep\nnn Will C M;1998;gr-qc/9811036
%gr-qc/9811036 
%From: Clifford M. Will <cmw@howdy.wustl.edu>
%The Confrontation between General Relativity and Experiment: A 1998 Update
%Authors: Clifford M. Will (Washington University, St. Louis)
%Comments: Lecture notes from the 1998 Slac Summer Institute on Particle Physics; 76 pages, 10
%figures

%%%%%%%%%%%%%%%%%%%%%%%%%%%%%%%%%%%%%%%%%%%%%%%%%%%%%%%%%%%%%%%%%%%%%

\bibitem{Weinberg72}
\rfbook\nn Weinberg S;1972;Gravitation and Cosmology;Wiley;{New York}

\bibitem{Eddington}
\rfbook\nn Eddington A;1920;Space, Time \& 
Gravitation;Cambridge University Press;Cambridge

\bibitem{HuGDM}
\rf\nn Hu W;1998;ApJ;506;485
%astro-ph/9801234
%Hu, Wayne


%%%%%%%%%%%%%%%%%%%%%%%%%%%%%%%%%%%%%%%%%%%%%%%%%%%%%%%%%%%%%%%%%%%%%
% CONSTRAINING THE EQUATION OF STATE
%%%%%%%%%%%%%%%%%%%%%%%%%%%%%%%%%%%%%%%%%%%%%%%%%%%%%%%%%%%%%%%%%%%%%

\bibitem{Garnavich98}
\rf\nnn Garnavich P M {\etal};1998;ApJ;509;74
% Garnavich, Peter M.; Jha, Saurabh;
% Challis, Peter; Clocchiatti, Alejandro;
% Diercks, Alan; Filippenko, Alexei V.;
% Gilliland, Ron L.; Hogan, Craig J.;
% Kirshner, Robert P.; Leibundgut, Bruno;
% Phillips, M. M.; Reiss, David;
% Riess, Adam G.; Schmidt, Brian P.;
% Schommer, Robert A.; Smith, R. Chris;
% Spyromilio, Jason; Stubbs, Chris;
% Suntzeff, Nicholas B.; Tonry, John;
% Carroll, Sean M.
% Supernova Limits on the Cosmic Equation of State

\bibitem{Starobinski98}
\rf\nnn Starobinsky A A;1998;JETP Lett.;68;757
% 757-763
% astro-ph/9810431
% Shows that g(z) determines H(z)/H_0 and derives simply 
% integral formula for how the reconstruction is done.

\bibitem{Efstathiou99}
\rf\nn Efstathiou G;1999;MNRAS;310;842
% Constraining the equation of state of the Universe from distant Type Ia supernovae and
% cosmic microwave background anisotropies

\bibitem{Perlmutter99}
\rf\nn Perlmutter S, \nnn Turner M S\multiand\nn White M;1999;Phys. Rev. Lett.;83;670
% Constraining dark energy with SN-Ia and large-scale structure

\bibitem{Cooray99}
\rf\nnn Cooray A R\dualand\nn Huterer D;1999;ApJL;513;L95
% astro-ph/9901097
% Title: Gravitational Lensing as a Probe of Quintessence
% Authors: Asantha R. Cooray, Dragan Huterer (University of Chicago)
% Comments: Accepted for publication in ApJ Letters (4 pages, including 4 figures)
% Journal-ref: Astrophys.J. 513 (1999) L95-L98

\bibitem{Huterer99}
\rf\nn Huterer D\dualand\nnn Turner M S;1999;Phys. Rev. D;60;081301
% astro-ph/9808133
% Title: Prospects for probing the dark energy via supernova distance measurements
% Authors: Dragan Huterer, Michael S. Turner (Chicago/Fermilab)
% Comments: 10 pages LaTeX with 3 eps figures. Reconstruction of the equation of state added + some minor changes. To appear
% in PRD Rapids

\bibitem{Huterer00}
\rfprep\nn Huterer D\dualand\nnn Turner M S;2001;64;123527
% astro-ph/0012510
% Title: Probing the dark energy: methods and strategies
% Authors: Dragan Huterer, Michael S. Turner
% Comments: 20 pages and 23 figures, revtex, submitted to Phys. Rev. D

\bibitem{Newman00}
\rf\nnn Newman J A;2000;ApJL;534;L11
% Newman, Jeffrey A.; Davis, Marc
% Measuring the Cosmic Equation of State with Counts of Galaxies

\bibitem{Lima00}
\rf\nnnn Lima J A S\dualand\nnn Alcaniz J S;2000;MNRAS;317;893
% Constraining the cosmic equation of state from old galaxies at high redshift

\bibitem{Saini00}
\rf\nnn Saini T D, \nn Raychaudhury S, \nn Sahni V\multiand\nnn Starobinsky A A;2000;Phys. Rev. Lett.;85;1162 
%astro-ph/9910231     Reconstructing the Cosmic Equation of State from Supernova distances
%Authors: Tarun Deep Saini (1), Somak Raychaudhury (1), Varun Sahni (1), A. A. Starobinsky (2,3)
%((1) IUCAA, Pune, (2) Landau Institute, Moscow, (3) MPIfA, Garching)
%Journal-ref: Phys.Rev.Lett. 85 (2000) 1162-1165
          
\bibitem{Bean01}
\rf\nn Bean R\dualand\nn Melchiorri A;2001;PRD;65;041302
% astro-ph/0110472
% Title: Current constraints on the dark energy equation of state
% Authors: Rachel Bean, Alessandro Melchiorri

\bibitem{Kujat01}
\rf\nn Kujat J, Linn A M, \nnn Scherrer R J\multiand Weinberg D H;2001;ApJ;572;1-14
% Title: Prospects for Determining the Equation of State of the Dark Energy: What can be Learned from Multiple Observables?
% Authors: Jens Kujat, Angela M. Linn, Robert J. Scherrer, David H. Weinberg

\bibitem{Maor02a}
\rf\nn Maor I, \nn Brustein R\multiand\nnn Steinhardt P J;2002;PRL;86;6
% Maor, Irit; Brustein, Ram; Steinhardt, Paul J.
% Limitations in Using Luminosity Distance to Determine the Equation of State of the Universe

\bibitem{Maor02b}
\rf\nn Maor I, \nn Brustein R, \nn McMahon J\multiand\nnn Steinhardt P J;2002;PRD;65;123003
% Maor, Irit; Brustein, Ram; McMahon, Jeff; Steinhardt, Paul J.
% Measuring the equation of state of the universe: Pitfalls and prospects

\bibitem{Weller02}
\rf\nn Weller J\dualand\nn Albrecht A;2002;PRD;65;103512
% astro-ph/0106079 [abs, ps, pdf, other] :
% Title: Future Supernovae observations as a probe of dark energy
% Authors: Jochen Weller (1), Andreas Albrecht (2) ((1) University of Cambridge, (2) UC Davis)

\bibitem{Wasserman02}
\rfprep\nn Wasserman I;2002;astro-ph/0203137
% Title: On the Degeneracy Inherent in Observational Determination of the Dark Energy Equation of State
% Authors: Ira Wassermsan
% astro-ph/0112221 [abs, ps, pdf, other] :

\bibitem{Spergel02}
\rfprep\nnn Spergel D N\dualand\nnn Starkman G D;2002;astro-ph/0204089
% Title: Using Supernovae to Determine the Equation of State of the Dark Energy: Is Shallow Better than Deep?
% Authors: David N. Spergel, Glenn D. Starkman

\bibitem{Peebles02}
\rfprep\nnnn Peebles P J E\dualand\nn Ratra B;2002;astro-ph/0207347
% Title: The Cosmological Constant and Dark Energy
% Authors: P. J. E. Peebles, Bharat Ratra
% Comments: 54 pages

\bibitem{Friemann02}
\rfprep\nnn Frieman J A, \nn Huterer D, \nnn Linder E V\multiand\nnn Turner M S;2002;astro-ph/0208100
% Title: Probing Dark Energy with Supernovae: Exploiting Complementarity with the Cosmic Microwave Background
% Authors: Joshua A. Frieman (Chicago/FNAL), Dragan Huterer (CWRU), Eric V. Linder (LBL), Michael S. Turner (Chicago/FNAL)
% Comments: 11 pages, 11 figures, submitted to Phys. Rev. D

\bibitem{Linder02}
\rfprep\nnn Linder E V\dualand\nn Huterer D;2002;astro-ph/0208138
% Title: Importance of Supernovae at z>1.5 to Probe Dark Energy
% Authors: Eric V. Linder (LBL), Dragan Huterer (CWRU)
% Comments: 6 pages, 7 figures, submitted to Phys. Rev. D
	  	  
\bibitem{Wang01}
\rf\nn Wang Y\dualand\nnn Garnavich P M;2001;ApJ;552;445
% astro-ph/0101040
% Title: Measuring Time-Dependence of Dark Energy Density from Type Ia Supernova Data
% Authors: Yun Wang, Peter M. Garnavich
% Comments: Minor changes for clarity. 21 pages including 7 figures. Submitted to ApJ on Aug. 2,
% 2000; resubmitted to ApJ
% Ouch! I discovered this 010118 and it partially scoops me! Nice paper.
      
\bibitem{pwindows}
\rfprep\nn Tegmark M\dualand\nn Zaldarriaga M;2002;astro-ph/0207047
% pwindows


%%% SECTION II:

\bibitem{Peebles93}
\rfbook\nnnn Peebles P J E;1993;Principles of Physical 
Cosmology;Princeton University Press;Princeton


\bibitem{Guerra00}
\rf\nnn Guerra E J, \nn Daly R\multiand\nn Wan L;2000;ApJ;544;659
%2000ApJ...544..659G
%Guerra, Erick J.; Daly, Ruth A.;
%Wan, Lin
%Global Cosmological Parameters Determined Using Classical
%Double Radio Galaxies

\bibitem{Pen97}
\rf\nn Pen U;1997;New. Astr.;2;309
% astro-ph/9610090
% Pen, UL.
% Measuring the universal deceleration using angular diameter distances to
% clusters of galaxies.
% NEW ASTRONOMY, 1997, V2 N4:309-317.
			      

%%%%%%%%%%%%%%%%%%%%%%%%%%%%%%%%%%%%%%%%%%%%%%%%%%%%%%%%%%%%%%%%%%%%%
%% Time-variation of fundamental constants:
%%%%%%%%%%%%%%%%%%%%%%%%%%%%%%%%%%%%%%%%%%%%%%%%%%%%%%%%%%%%%%%%%%%%%

\bibitem{Damour96}
\rf\nn Damour T\dualand\nn Dyson F;1996;Nucl. Phys. B;480;37
% |alphadot/alpha| < 7e-17 (95%).
      
\bibitem{Webb99}
\rf\nnn Webb J K, \nnn Flambaum V V, \nnn Churchill C W,
\nnn Drinkwater M J\multiand\nnn Barrow J D;1999;Phys. Rev. Lett.;82;884
% Webb, John K.;
% Flambaum, Victor V.;
% Churchill, Christopher W.;
% Drinkwater, Michael J.;
% Barrow, John D.
% Search for Time Variation of the Fine Structure Constant
    
\bibitem{Dickey94}
\rf\nnn Dickey J O {\etal};1994;Science;265;482
% Limits on time-variation of G

\bibitem{Williams96}
\rf\nnn Williams J G, 
\nnn Newhall X X\multiand\nnn Dickey J O;1996;Phys. Rev. D;53;6730
% Limits on time-variation of G: Gdot/G < 6e-12/year 
      


%%%%%%%%%%%%%%%%%%%%%%%%%%%%%%%%%%%%%%%%%%%%%%%%%%%%%%%%%%%%%%%%%%%%%
%% Alcock-Paczynski test:
%%%%%%%%%%%%%%%%%%%%%%%%%%%%%%%%%%%%%%%%%%%%%%%%%%%%%%%%%%%%%%%%%%%%%

\bibitem{AlcockPaczynski79}
\rf\nn Alcock C\dualand\nn Paczynski B;1979;Nature;281;358
% An evolution free test for non-zero cosmological constant

\bibitem{Nair99}
\rf\nn Nair V;1999;ApJ;522;569
%Nair, Vibhat
%Geometric Distortion of the Correlation Function of Lyman-Break Galaxies 



%%%%%%%%%%%%%%%%%%%
\bibitem{parameters2}
\rf\nnn Eisenstein D J, 
\nn Hu W\multiand Tegmark M;1999;ApJ;518;2
% astro-ph/9807130



%%%%%%%%%%%%%%%%%%%%%%%%%%%%%%%%%%%%%%%%%%%%%%%%%%%%%%%%%%%%%%%%%%%%%
%% SN 1a data refs:
%%%%%%%%%%%%%%%%%%%%%%%%%%%%%%%%%%%%%%%%%%%%%%%%%%%%%%%%%%%%%%%%%%%%%

\bibitem{Perlmutter98}
\rf\nn Perlmutter S {\etal};1998;Nature;391;51

\bibitem{Riess98}
\rf\nnn Riess A G {\etal};1998;Astron. J.;116;1009
% Observational Evidence from Supernovae for an Accelerating Universe and a Cosmological Constant% astro-ph/9805201
% The Astronomical Journal, Volume 116, Issue 3, pp. 1009-1038.

%%%%%%%%%%%%%%%%%%%%%%%%%%%%%%%%%%%%%%%%%%%%%%%%%%%%%%%%%%%%%%%%%%%%%
%% Angular diameter distance degeneracy:
%%%%%%%%%%%%%%%%%%%%%%%%%%%%%%%%%%%%%%%%%%%%%%%%%%%%%%%%%%%%%%%%%%%%%

\bibitem{WhiteComplementarity}
\rf\nn White M;1998;ApJ;506;495
% astro-ph/9802295
% Complementary Measures of the Mass Density and Cosmological Constant
% Astrophys.J. 506 (1998) 495

\bibitem{complementarity}
\rfprep\nn Tegmark M, \nnn Eisenstein D J, \nn Hu W\multiand\nn Kron R;1998;astro-ph/9805117

\bibitem{Huey99}
\rf\nn Huey G, \nn Wang L, \nn Dave R, \nnn Caldwell R R\multiand\nnn Steinhardt P J;1999;Phys. Rev. D;59;063005
%astro-ph/9804285 [abs, src, ps, other] :
%Title: Resolving the Cosmological Missing Energy Problem
%Authors: Greg Huey, Limin Wang, R. Dave, R. R. Caldwell, Paul J. Steinhardt
%Comments: 6 pages, Latex, 4 postscript figures; revised analysis to include gravitational lensing
%Journal-ref: Phys.Rev. D59 (1999) 063005
    
%%%%%%%%%%%%%%%%%%%%%%%%%%%%%%%%%%%%%%%%%%%%%%%%%%%%%%%%%%%%%%%%%%%%%
\bibitem{Wald84}
\rfbook\nnn Wald R M;1984;General 
Relativity;University of Chicago Press;Chicago

\bibitem{Fisher35}
\rf\nnn Fisher R A;1935;J. Roy. Stat. Soc.;98;39

\bibitem{karhunen}
\rf\nn Tegmark M, \nnn Taylor A N\multiand\nnn Heavens A F;1997;ApJ;480;22

%\bibitem{Colless98}
%\rfprep\nn Colless M;1998;astro-ph/9804079
%%Title: First results from the 2dF galaxy redshift survey
%%Authors: Matthew Colless (MSSSO)
%%Comments: To appear in Phil.Trans.R.Soc.Lond.A, 12 pages, 8 figures, LaTeX and rspublic.sty
%% There still is no ``official'' 2dF galaxy redshift survey ref.
%
%\bibitem{York00}
%\rf\nnn York D G;2000;Astron. J.;120;1579
%% astro-ph/0006396
%% Title: The Sloan Digital Sky Survey: Technical Summary
%% Authors: D.G. York, et al. (The SDSS Collaboration)
%% Comments: 9 pages, 7 figures, AAS Latex. To appear in AJ, Sept 2000
%
%%%%%%%%%%%%%%%%%%%%%%%%%%%%%%%%%%%%%%%%%%%%%%%%%%%%%%%%%%%%%%%%%%%%%%
%% Redshift space distortions:
%%%%%%%%%%%%%%%%%%%%%%%%%%%%%%%%%%%%%%%%%%%%%%%%%%%%%%%%%%%%%%%%%%%%%
%\bibitem{Kaiser87}
%\rf Kaiser N;1987;MNRAS;227;1
%
%\bibitem{HamiltonReview}
%\rfprep\nnnn Hamilton A J S;1997;astro-ph/9708102 
%% Title: Linear Redshift Distortions: A Review
%% Author: A. J. S. Hamilton (JILA, U. Colorado)
%% Comments: 91 pages, including 8 embedded EPS figures. LaTeX, crckapb.sty. Final version of invited review to appear in
%% Hamilton, D. (ed.) Ringberg Workshop on Large-Scale Structure, held at Ringberg Castle, Germany, 23-28 September 1996,
%% Kluwer Academic, Dordrecht. Numerous minor revisions, some errors fixed. One major revision: careful discussion of difference
%% between real and redshift selection function
%
%%%%%%%%%%%%%%%%%%%%%%%%%%%%%%%%%%%%%%%%%%%%%%%%%%%%%%%%%%%%%%%%%%%%%%
%%% Higher order moments:
%%%%%%%%%%%%%%%%%%%%%%%%%%%%%%%%%%%%%%%%%%%%%%%%%%%%%%%%%%%%%%%%%%%%%%
%
%\bibitem{Szapudi00}
%\rfprep\nn Szapudi I, \nn Postman M, \nnn Lauer T R\multiand\nn Oegerle W;2000;astro-ph/0008131
%% Title: Observational Constraints on Higher Order Clustering up to $z\simeq 1
%% Authors: István Szapudi (CITA), Marc Postman (STScI), Tod R. Lauer (NOAO), William
%% Oegerle (Johns Hopkins)
%% Comments: 28 pages, 4 figures included, ApJ, accepted, minor changes
%
%\bibitem{Feldman00}
%\rfprep\nnn Feldman H A, \nnn Frieman J A, \nnn Fry J N\multiand\nn Scoccimarro R;2000;astro-ph/0010205
%% Title: Constraints on Galaxy Bias, Omega_m, and Primordial Non-Gaussianity from the PSCz
%% Survey Bispectrum
%% Authors: Hume A. Feldman (Kansas) Joshua A. Frieman (Fermilab), J. N. Fry (Florida), Roman
%% Scoccimarro (IAS)
%
%%%%%%%%%%%%%%%%%%%%%%%%%%%%%%%%%%%%%%%%%%%%%%%%%%%%%%%%%%%%%%%%%%%%%%
%
%\bibitem{Davis00}
%\rfprep\nn Davis M\dualand\nn Newman J;2000;astro-ph/0012189
%%The DEEP2 Redshift Survey
%%Authors: Marc Davis, Jeffrey Newman (UC Berkeley), Sandra Faber, Andrew Phillips (UC Santa Cruz)
%%Comments: 6 pages, to appear in Proc. of the ESO/ECF/STSCI workshop on Deep Fields, Garching Oct
%%2000, (Publ: Springer)
%
%\bibitem{LeFevre01}
%\rfprep\nn {Le F\`evre} O {\etal};2001;astro-ph/0101034
%% The VIRMOS-VLT Deep Survey
%% Authors: O. Le Fèvre, G. Vettolani, D. Maccagni, D. Mancini, A. Mazure, Y. Mellier, J.P. Picat, M.
%% Arnaboldi, S. Bardelli, E. Bertin, G. Busarello, A. Cappi, S. Charlot, G. Chincarini, S. Colombi, B.
%% Garilli, L. Guzzo, A. Iovino, V. Le Brun, M. Longhetti, G. Mathez, P. Merluzzi, H.J. McCracken, R.
%% Pellò, L. Pozzetti, M. Radovich, V. Ripepi, P. Saracco, R. Scaramella, M. Scodeggio, L. Tresse, G.
%% Zamorani, E. Zucca
%% Comments: 5 pages including figures; to appear in Proc. of the ESO/ECF/STSCI "Deep Fields"
%% workshop, Garching Oct 2000, (Publ: Springer)
%
%\bibitem{Yamamoto01}
%\rfprep\%nn Yamamoto K\dualand\nn Nishioka H;2001;astro-ph/astro-ph/0101172 
% Can Ge%ometric Test Probe the Cosmic Equation of State ?
% Author%s: Kazuhiro Yamamoto, Hiroaki Nishioka (Hiroshima University)
% Commen%ts: 11 pages, including 3 figures, accepted for publication in ApJL
% Feasib%ility of the geometric test as a probe of the cosmic equation of state of the dark
% energy% is discussed assuming the future 2dF QSO sample.
%
%%%%%%%%%%%%%%%%%%%%%%%%%%%%%%%%%%%%%%%%%%%%%%%%%%%%%%%%%%%%%%%%%%%%%%
%%% LyA P(k):
%%%%%%%%%%%%%%%%%%%%%%%%%%%%%%%%%%%%%%%%%%%%%%%%%%%%%%%%%%%%%%%%%%%%%%
%
%\bibitem{Croft99}
%\rf\nnnn Croft R A C, \nnn Weinberg D H, \nn Pettini M,
%\nn Hernquist L\multiand\nn Katz N;1999;ApJ;520;1
%% Croft, Rupert A. C.;
%% Weinberg, David H.;
%% Pettini, Max; Hernquist, Lars;
%% Katz, Neal
%% Cosmological Limits on the Neutrino Mass from the LyAlpha Forest
%
%\bibitem{WhiteCroft00}
%\rf\nn White M\dualand\nnnn Croft R A C;2000;ApJ;539;497
%% astro-ph/0001247
%% Title: Suppressing Linear Power on Dwarf Galaxy Halo Scales
%% Authors: Martin White, Rupert A.C. Croft
%% Comments: 9 pages, 9 figures
%% Journal-ref: Astrophys.J. 539 (2000) 497
%    
%\bibitem{McDonald00}
%\rf\nn McDonald P, \nn {Miralda-Escud\'e} J, \nn Rauch M, \nnnn Sargent W L W, \nnn Barlow T A,
%\nn Cen R\multiand\nnn Ostriker J P;2000;ApJ;543;1
%% McDonald, Patrick;
%% Miralda-Escudé, Jordi;
%% Rauch, Michael;
%% Sargent, Wallace L. W.;
%% Barlow, Tom A.; 
%% Cen, Renyue;
%% Ostriker, Jeremiah P.
%% The Observed Probability Distribution Function, Power
%% Spectrum, and Correlation Function of the Transmitted Flux in
%% the LY&alpha; Forest
%
%\bibitem{LyA}
%\rfprep\nn Zaldarriaga M, \nn Hui L\multiand\nn Tegmark M;2000;astro-ph/0011559
%% Title: Constraints from the Lyman alpha forest power spectrum
%% Authors: Matias Zaldarriaga, Lam Hui, Max Tegmark
%    
%\bibitem{Croft00}
%\rfprep\nnnn Croft R A C, \nnn Weinberg D H, \nn Bolte M,
%\nn Burles S, \nn Hernquist L, \nn Katz N, \nn Kirkman D\multiand\nn Tytler D;2000;astro-ph/0012324
%% Title: Towards a Precise Measurement of Matter Clustering: Lyman-alpha Forest Data at
%% Redshifts 2-4
%% Authors: Rupert A.C. Croft (CfA), David H. Weinberg (Ohio State), Mike Bolte (UCO/Lick),
%% Scott Burles (Fermilab), Lars Hernquist (CfA), Neal Katz (UMass), David Kirkman (UCSD),
%% David Tytler (UCSD)
%
%%%%%%%%%%%%%%%%%%%%%%%%%%%%%%%%%%%%%%%%%%%%%%%%%%%%%%%%%%%%%%%%%%%%%%
%%% COSMIC SHEAR:
%%%%%%%%%%%%%%%%%%%%%%%%%%%%%%%%%%%%%%%%%%%%%%%%%%%%%%%%%%%%%%%%%%%%%%
%\bibitem{Wittman00}
%\rf\nnn Wittman D M, \nn Tyson J A, \nn Kirkman D, 
%\nn {Dell'Antonio} I\multiand\nn Bernstein G;2000;Nature;405;143
%% Wittman, David M.;
%% Tyson, J. Anthony;
%% Kirkman, David;
%% Dell'Antonio, Ian;
%% Bernstein, Gary
%% Detection of weak gravitational lensing distortions of distant
%% galaxies by cosmic dark matter at large scales
%
%\bibitem{Waerbeke00}
%\rf\nn {Van Waerbeke} L {\etal};2000;A\&A;358;30
%% Van Waerbeke, L.; Mellier, Y.;
%% Erben, T.; Cuillandre, J. C.;
%% Bernardeau, F.; Maoli, R.;
%% Bertin, E.; Mc Cracken, H. J.;
%% Le Fèvre, O.; Fort, B.;
%% Dantel-Fort, M.; Jain, B.;
%% Schneider, P.
%% Detection of correlated galaxy ellipticities from CFHT data: first
%% evidence for gravitational lensing by large-scale structures
%
%\bibitem{Bacon00}
%\rf\nnn Bacon D J, \nnn Refregier A R\multiand\nnn Ellis R S;2000;MNRAS;318;625
%% Detection of weak gravitational lensing by large-scale structure
%
%\bibitem{Kaiser00}
%\rfprep\nn Kaiser N, \nn Wilson G\multiand\nnn Luppino G A;2000;astro-ph/0003338
%% Title: Large-Scale Cosmic Shear Measurements
%% Authors: Nick Kaiser, Gillian Wilson, Gerard A. Luppino
%% Comments: submitted to ApJ Letters
%
%\bibitem{Rhodes01}
%\rfprep\nn Rhodes J, Refregier A\multiand\nnn Groth E J;2001;astro-ph/0101213
%%Title: Detection of Cosmic Shear with the HST Survey Strip
%%Authors: J. Rhodes (1 and 2), A. Refregier (3), E. J. Groth (2) ((1) Goddard Space Flight Center,
%%(2) Princeton University, (3) Cambridge University)
%%Comments: 4 pages, 2 figures
%
%%%%%%%%%%%%%%%%%%%%%%%%%%%%%%%%%%%%%%%%%%%%%%%%%%%%%%%%%%%%%%%%%%%%%%
%\bibitem{HuTomography}
%\rf\nn Hu W;1999;ApJL;522;L21
%% Hu, Wayne
%% Power Spectrum Tomography with Weak Lensing 
%% astro-ph/9904153
%
%\bibitem{Fischer00}
%\rf\nn Fischer P;2000;Astron. J.;120;1198
%% Fischer, Philippe;
%% Weak Lensing with Sloan Digital Sky Survey Commissioning Data:
%% The Galaxy-Mass Correlation Function to 1 H-1 Mpc
%
%\bibitem{Moessner98}
%\rf\nn Moessner R\dualand\nn Jain B;1998;MNRAS;294;L18
%% Moessner, R.; Jain, Bhuvnesh
%% Angular cross-correlation of galaxies - A probe of gravitational
%% lensing by large-scale structure
%
%
%
%\bibitem{Zalda99}
%\rfprep\nn Zaldarriaga M;1999;astro-ph/9910498
%% Title: Lensing of the CMB: Non Gaussian aspects
%% Authors: Matias Zaldarriaga
%% Comments: Changes to match accepted version in PRD, 20 pages 10 figures. Better resolution
%% images of the figures can be found at this http URL
%
%\bibitem{Guzik99}
%\rfprep\nn Guzik J, \nn Seljak U\multiand\nn Zaldarriaga M;1999;astro-ph/9912505
%% Title: Lensing effect on polarization in microwave background: extracting convergence power
%% spectrum
%% Authors: Jacek Guzik (Jagiellonian University), Uros Seljak (Princeton University), Matias
%% Zaldarriaga (IAS)
%% Comments: 17 pages, 5 figures, accepted for publication in PRD
%    
%\bibitem{Suginohara98}
%\rf\nn Suginohara M, \nn Suginohara T\multiand\nnn Spergel D N;1998;ApJ;495;511 
%% astro-ph/9705134
%% Title: Cross-Correlating Cosmic Microwave Background Radiation Fluctuations with Redshift
%% Surveys: Detecting the Signature of Gravitational Lensing
%% Authors: Maki Suginohara, Tatsushi Suginohara, David N. Spergel (Princeton University)
%% Comments: 13 pages, 4 postscript figures included; Uses aaspp4.sty (AASTeX v4.0); Accepted for
%% publication in Astrophysical Journal, Part 1
%% Journal-ref: Astrophys.J. 495 (1998) 511 
%     
%\bibitem{Mohr00}
%\rfprep\nnn Mohr J J, \nn Haiman Z\multiand\nnn Holder G P;2000;astro-ph/0004244
%% Title: Galaxy Cluster Baryon Fractions, Cluster Surveys and Cosmology
%% Authors: Joseph J. Mohr, Zoltan Haiman, Gilbert P. Holder
%% Comments: 10 pages; Plenary contribution to PASCOS99, December 1999
%
%
%%%%%%%%%%%%%%%%%%%%%%%%%%%%%%%%%%%%%%%%%%%%%%%%%%%%%%%%%%%%%%%%%%%%%%
%%% CLUSTER ABUNDANCE:
%%%%%%%%%%%%%%%%%%%%%%%%%%%%%%%%%%%%%%%%%%%%%%%%%%%%%%%%%%%%%%%%%%%%%%
%
%
%\bibitem{Bahcall98}
%\rf\nnn Bahcall N A\dualand\nn Fan X;1998;ApJ;504;1
%% astro-ph/9803277
%% Title: The Most Massive Distant Clusters: Determining Omega and sigma_8
%% Authors: Neta A. Bahcall, Xiaohui Fan (Princeton University Observatory)
%% Comments: 14 pages, 4 Postscript figures, ApJ in press
%% Journal-ref: ApJ, 504 (1998), 1
%    
%\bibitem{Eke98}
%\rf\nnn Eke V R, \nn Cole S, \nnn Frenk C S\multiand\nnn Henry J P;1998;MNRAS;298;1145
%% astro-ph/9802350
%% Measuring Omega_0 using cluster evolution
%% Authors: V. R. Eke, S. Cole, C. S. Frenk, J. P. Henry
%% Comments: 17 pages, 15 figures, submitted to MNRAS
%% The evolution of the galaxy cluster abundance depends sensitively on the value of the cosmological density parameter, Omega_0. Recent
%% ASCA data are used to quantify this evolution as measured by the X-ray temperature function. A chi^2 minimization fit to the cumulative
%% temperature function, as well as a maximum likelihood estimate (which requires additional assumptions about cluster luminosities), lead to
%% the estimate Omega_0 \approx 0.45+/-0.2 (1-sigma statistical error). Various systematic uncertainties are considered, none of which
%% enhance significantly the probability that Omega_0=1. These conclusions hold for models with or without a cosmological constant. The
%% statistical uncertainties are at least as large as the individual systematic errors that have been considered here, suggesting that additional
%% temperature measurements of distant clusters will allow an improvement in this estimate. An alternative method that uses the highest
%% redshift clusters to place an upper limit on Omega_0 is also presented and tentatively applied, with the result that Omega_0=1 can be ruled
%% out at the 98 per cent confidence level. Whilst this method does not require a well-defined statistical sample of distant clusters, there are
%% still modeling uncertainties that preclude a firmer conclusion at this time.
%
%\bibitem{Henry00}
%\rf\nnn Henry J P;2000;ApJ;534;565
%% Omega_m = 0.44+/-.12 for flat models from evolution of cluster
%% X-ray temperature function

\end{references}
\end{document}